\documentstyle[11pt,aaspp4,flushrt]{article}

\def\mi{\ifmmode\overline{m}_I\else$\overline{m}_I$\fi}
\def\mv{\ifmmode\overline{m}_V\else$\overline{m}_V$\fi}
\def\Mi{\ifmmode\overline{M}_I\else$\overline{M}_I$\fi}
\def\mim{\ifmmode\overline{m}_I^0\else$\overline{m}_I^0$\fi}
\def\dmod{\ifmmode(m{-}M)\else$(m{-}M)$\fi}
\def\vi{\ifmmode(V{-}I)\else$(V{-}I)$\fi}
\def\viz{\ifmmode(V{-}I)_0\else$(V{-}I)_0$\fi}
\def\dn{\ifmmode D_n{-}\sigma\else$ D_n{-}\sigma$\fi}
\def\avemi{\ifmmode\langle\overline{m}_I^0\rangle\else$\langle\overline{m}_I^0\rangle$\fi}

\def\itemitem{\par\indent \hangindent2\parindent \textindent}

\lefthead{Tonry, Blakeslee, Ajhar and Dressler}
\righthead{SBF Survey. I.}

\begin{document}

\title{The SBF Survey of Galaxy Distances. I.}
\title{Sample Selection, Photometric Calibration, and the Hubble Constant\altaffilmark{1}}

\author{John L. Tonry\altaffilmark{2} and John P. Blakeslee\altaffilmark{2}}
\affil{Physics Dept. Room 6-204, MIT, Cambridge, MA 02139;}
\authoremail{jt@antares.mit.edu}
\authoremail{john@arneb.mit.edu}

\author{Edward A. Ajhar\altaffilmark{2}}
\affil{Kitt Peak National Observatory, National Optical Astronomy
Observatories, P.O. Box 26732 Tucson, AZ 85726;}
\authoremail{ajhar@noao.edu}

\author{Alan Dressler}
\affil{Carnegie Observatories, 813 Santa Barbara St., Pasadena, CA 91101}
\authoremail{dressler@omega.ociw.edu}

\altaffiltext{1}{Observations in part from the Michigan-Dartmouth-MIT 
(MDM) Observatory.}
\altaffiltext{2}{Guest observers at the Cerro Tololo Interamerican
Observatory and the Kitt Peak National Observatory, National Optical
Astronomy Observatories, which are operated by AURA, Inc., under
cooperative agreement with the National Science Foundation.}

\begin{abstract}

We describe a program of surface brightness fluctuation (SBF)
measurements for determining galaxy distances.  This paper presents the
photometric calibration of our sample and of SBF in general.  Basing
our zero point on observations of Cepheid variable stars we find that
the absolute SBF magnitude in the Kron-Cousins $I$ band correlates well
with the mean \viz\ color of a galaxy according to 
$$ \overline M_I = (-1.74\pm0.07)\, +\, (4.5\pm0.25)\, [ \viz - 1.15 ]$$
for $1.0<\vi<1.3$.  This agrees well with
theoretical estimates from stellar population models.

\noindent Comparisons between SBF distances and a variety of other
estimators, including Cepheid variable stars, the Planetary Nebula
Luminosity Function (PNLF), Tully-Fisher (TF), \dn, SNII, and SNIa,
demonstrate that the calibration of SBF is universally valid and that
SBF error estimates are accurate.  The zero point given by Cepheids,
PNLF, TF (both calibrated using Cepheids), and SNII is in units of Mpc;
the zero point given by TF (referenced to a distant frame), \dn, and
SNIa is in terms of a Hubble expansion velocity expressed in km/s.
Tying together these two zero points yields a Hubble constant of 
$$H_0 = 81\pm6\;\hbox{km/s/Mpc}.$$ 
As part of this analysis, we present SBF distances to 12 nearby groups
of galaxies where Cepheids, SNII, and SNIa have been observed.

\end{abstract}


\section{Introduction and Sample Selection}

\subsection{Background}

The surface brightness fluctuation (SBF) method of distance
determination works by measuring the ratio of the second and 
first moments of the stellar luminosity function in a galaxy.
This ratio, called $\overline L$, is then the luminosity-weighted,
average luminosity of a stellar population and is roughly equal
to the luminosity of a single giant star.  In terms of magnitudes,
this quantity is represented as $\overline M$,
the absolute ``fluctuation magnitude.''
What we measure, of course, is the apparent fluctuation magnitude
in a particular photometric band, in our case the $I$~band,
\mi.  In order to be useful as a distance estimator,
\mi\ must be calibrated, either empirically, by tying 
the measurements
to the Cepheid distance scale, or theoretically, according
to stellar population synthesis models.

Tonry and Schneider (1988) were the first to quantify the SBF phenomenon.
Their method was based on a measurement of the amount of power on the
scale of the point spread function in the power spectrum of a CCD image.
They applied this method to images of the galaxies M32 and NGC 3379.
Subsequent work by Tonry, Luppino, and Schneider (1988) and
Tonry, Ajhar, \& Luppino (1989, 1990) revised and refined the analysis
techniques and presented further observations in $V$, $R$, 
and $I$ of early-type galaxies
in Virgo, Leo, and the Local Group.
Tonry et al.\ (1990) found that the $I$~band was most 
suitable for measuring distances and attempted to calibrate 
the SBF method theoretically using the Revised Yale Isochrones 
(Green, Demarque, \& King 1987).
There were obvious problems with this calibration, however, so
a completely empirical calibration for \Mi\ was presented
by Tonry (1991).  The calibration was based on the variation of 
\mi\ with \viz\ color in the Fornax cluster and took
its zero point from the Cepheid distance to M31.  Tonry used this
calibration to derive the Hubble constant.
A detailed review of the modern SBF technique can be found 
in Jacoby et al.\ (1992), which also provides some historical
context for the method.

In recent years, the SBF technique has been used to measure distances
and study a variety of stellar populations in several different bands.
$K$-band SBF observations have been reported by Luppino \& Tonry (1993),
Pahre \& Mould (1994), and Jensen, Luppino, \& Tonry (1996).  These
studies find that $\overline m_K$ is also a very good distance estimator.
Dressler (1993) has measured $I$-band SBF in Centaurus ellipticals,
finding evidence in support of the Great Attractor model.
Lorenz et al.\ (1993) have measured $I$-band SBF in Fornax, and
Simard \& Pritchet (1994) have reported
distances to two Coma~I galaxies using $V$-band SBF observations.
Ajhar \& Tonry (1994) reported measurements of \mi\ and
$\overline m_V$ for 19 Milky Way globular clusters and considered
the implications for both the distance scale and stellar populations.
Tiede, Frogel, \& Terndrup (1995) measured \mi\ and
$\overline m_V$ for the bulge of the Milky Way and derived the
distance to the Galactic center.
Sodemann \& Thomsen (1995, 1996) have used fluctuation colors and
radial gradients to investigate stellar populations in galaxies.
Finally, an enormous amount of progress has been achieved on the theoretical
SBF front through the stellar population models of Worthey (1993a,b, 1994),
Buzzoni (1993, 1995), and Yi (1996).

\subsection{Genesis of the SBF Survey}

When it became apparent that $I$-band SBF observations could indeed
provide accurate and reliable distances to galaxies, we undertook a
large survey of nearby galaxies.  The sample selection is not
precisely defined because the measurement of SBF depends on a number
of criteria which are not ordinarily cataloged, such as dust content.
In addition, because the measurement of SBF is fairly expensive in
terms of telescope time and quality of seeing, it simply was not
possible to observe all early-type galaxies within some magnitude
limit out to a redshift which is large enough to make peculiar
velocities negligible.  Nevertheless, we have tried to manage fairly
complete coverage of early-type galaxies within 2000~km/s and brighter
than $B = 13.5$, and we have significant coverage beyond those limits.

Comparison with
the Third Reference Catalog of Bright Galaxies 
(de Vaucouleurs et al.\ 1991)
reveals that of the early-type galaxies ($T<0$) with $B \le 13.5$ in
the RC3, we have observed 76\% with heliocentric velocity
$v < 1000$~km/s, 73\% with $1000 < v < 1400$, 
64\% with $1400 < v < 2000$, 49\% with $2000 < v < 2800$, and we have
data for more than 40 galaxies with $v > 2800$~km/s.  Virtually all of the 
galaxies closer than $v<2000$ where we lack data are S0s
for which measuring SBF is complicated by
dust and/or disk/bulge problems, and since many of them are in the cores of
clusters such as Virgo, we do not regard their distances as being
important enough to delay completion of our survey.
The survey is, however, an ongoing project, with some data still to be reduced.
About 50\% of our sample is listed as E galaxies ($T\le-4$), about
40\% as S0s, and 10\% as ``spirals'' ($T\ge0$).
Our sample of galaxies is drawn from the entire sky, and the
completeness was mainly driven by the vicissitudes of weather and
telescope time, so the sampling is fairly random.  The survey
includes a large number of galaxies in the vicinity of the Virgo
supercluster, and the next paper in this series will present an analysis
of their peculiar motions.  

The following section describes the SBF survey in more detail,
including the observations, photometric reductions, and consistency checks.
In Section 3 we use our observations of galaxies in groups to derive
the dependence of \Mi\ on \viz.  Seven of these groups 
also have Cepheid distances, which we use to set the zero point of the
\Mi--\vi\ relation.  This new \Mi\ calibration
agrees well with theory and supersedes the old calibration of Tonry
(1991).  We then compare our distances
to those found using a number of other methods.  In Section 4 we
discuss the tie to the large-scale Hubble flow and implications
for the value of $H_0$.  The final section provides a summary of our
main conclusions.

\section{Observations and Reductions}

All in all, the SBF survey extends over some 40 observing runs at 6
telescopes.  Table 1 lists these runs along with some salient
information.  Note that the date of the run is coded in the name
as (Observatory)MMYY; the remaining columns are described below.

The normal observing procedure when the skies were clear was to
obtain sky flats each night and observe a number of Landolt standard
stars.  We preferred observing the faint standard star fields of 
Landolt (1992) in which there are several stars per CCD field and
where the observations are long enough that shutter timing is not a
problem.  Table 2 gives our usual fields.
During a typical night
we would observe about 10 fields comprising perhaps 50 stars over a
range of airmasses from 1.1 to 2.0.  We also strived to observe stars
over a wide
range of color ranging from $0 < \vi < 2$.  Because there is substantial
fringing seen in the $I$ band with thin CCDs, at some point in a
run we would spend several hours looking at a blank field in order to
build up a ``fringe frame''.  We have found the fringing pattern
for a given CCD and filter (although not the amplitude) to be
remarkably stable from night to night (even year to year).  Hence, a
single fringe frame was used to correct an entire run's data, and we
usually collected a new one for each run.

The reductions of the photometry proceed by bias subtraction,
flattening, and following Landolt (1992), 
summing the net flux from photometric standards within a
14\arcsec\ aperture.  We also estimate a flux error from the sky brightness
and variability over the image and remove any stars whose expected
error is greater than 0.02 magnitude.  Once all the photometric
observations from a run have been reduced, 
we fit the results according to 
\begin{equation}
m = m_1 - 2.5\log(f) - A\;\sec z + C\;\vi,
\end{equation}
where $f$ is the flux from the star in terms of electrons per second.
We have found that $m_1$ and $C$ are constant during a run
with a given CCD and filter, so we fit for a single value for these
parameters and extinction coefficients $A$ for each night.  The rms
residual of the fit is typically about 0.01 magnitude which is 
satisfactory accuracy for our purposes.  Table 1 lists typical values
for $m_1$, $A$, and $C$ for each run in the two filters $V$ and $I$.
Note the havoc in the extinction caused by the eruption by Pinatubo in 1991.

Galaxy reductions proceed by first bias subtraction,
division by a flat field, and subtraction of
any fringing present in $I$ band data.
We always take multiple images of a galaxy with the telescope
moved by several arcseconds between images, and determine these offsets
to the nearest pixel.  Any bad pixels or columns are
masked out, and the images are shifted into registration.  We next
run a program called ``autoclean'' which identifies cosmic rays in
the stack of images and removes them by replacement with appropriately
scaled data from the rest of the stack.  Autoclean also gives us
an estimate of how photometric the sequence of observations was by
producing accurate flux ratios between the exposures.
Finally we make a mask
of the obvious stars and companion galaxies in the cleaned image and
determine the sky background by fitting the outer parts of the galaxy
image with an $r^{1/4}$ profile plus sky level.  This is usually done 
simultaneously for $V$ and $I$ images, and when the sky levels are
determined, we also compute \viz\ colors as a function of radius
from the center of the galaxy.

In order to knit all of our observations into a consistent photometric
system, we attempted to make sure that there were overlap
observations between runs, and we developed a pair of programs called
``apphot'' and ``apcomp'' to compare observations.  ``Apphot'' 
converts a galaxy image with its photometric calibration into a 
table of circular aperture photometry.  This only depends on
plate scale (which is well known) and therefore permits comparison of
different images regardless of their angular orientation.  ``Apcomp''
then takes the aperture photometry from two observations and fits the
two profiles to one another using a photometry scale offset and a
relative sky level.  These two programs can give us accurate offsets
between the photometry of two images, good at the 0.005 magnitude
level.

We learned, however, that good seeing is much more common than
photometric weather, and we realized that many of our
``photometric'' observations were not reliable at the 0.01 magnitude
level.  As the survey progressed and the number of overlaps increased,
we also realized that although we only need 0.05 magnitude
photometry of \mi, \Mi\ is sensitive enough to
\viz\ color that we needed better photometry.  The existing 
photoelectric (PE) photometry, although probably very good in the
mean, is neither extensive enough nor accurate enough to serve to
calibrate the survey.

We also became aware that there are many peculiar CCD and shutter
effects which make good photometry difficult.  For example, we
have found photometry with Tektronix (SITe) CCDs particularly 
challenging for reasons we do not fully understand.  Because of their
high quantum efficiency and low noise they have been the detectors
of choice, but run to run comparisons with apphot and apcomp 
show consistent zero point offsets at the 0.05 magnitude level.  While
not a serious problem for \mi, we had to do much better in measuring
\viz.

Accordingly, we undertook an auxiliary survey in 1995 of a substantial
fraction of our SBF survey from the McGraw-Hill 1.3-m telescope at
the MDM Observatory.  We
shared the time with another program and used only nights which were
photometric, as judged by the observer at the time and
as revealed later from the quality of the photometric standard
observations.  We did not use Tek CCDs but primarily used the 
front-illuminated, Loral
2048$^2$ CCD Wilbur (Metzger et al. 1993),
we used filters which match $V$ and $I_{KC}$ as 
closely as possible, and the large field of view permitted us to make
very good estimates of sky levels.  Over 5 runs this comprised about
600 observations in $V$ and $I$.  We made certain to have a generous
overlap between these observations and all our other observing runs,
reaching well south to tie to the CTIO and LCO data.

We then performed a grand intercomparison of all the photometric
data in order to determine photometric offsets from run to run.
Using apphot and apcomp, we determined offsets between
observations, and we built up a large table of comparison pairs.
In addition, photoelectric (PE) photometry from deVaucouleurs
and Longo (1988), Poulain and Nieto (1994), and Buta and Williams (1995)
served as additional sources of comparison, and we
computed differences between PE and our photometry for every galaxy in
common.  We have found that \viz\ colors often show somewhat better
agreement than the individual $V$ and $I$ measurements, presumably
because thin clouds are reasonably gray, so we also compared colors
directly in addition to photometric zero points.

The results are illustrated in Figure 1.
In each of three quantities $V$, $I$, and \viz, we fitted for 
zero point offsets for each run (photoelectric sources were 
considered to be a ``run''), minimizing the pairwise differences.
We set the overall zero point by insisting that the median run
offset be zero.  Upon completion, we found that the rms
of the zero point offsets to be 0.029 mag, and the rms scatter
of individual comparisons between CCD data to be 0.030 mag in
$V$, 0.026 mag in $I$, and 0.024 mag in \viz.  The scatter was 
bigger for CCD-PE zero point comparisons, 0.047 mag in both $V$ and
$I$.  The ``zero point offsets'' for the photoelectric
photometry were 0.003 mag in $V$, 0.017 mag in $I$, and 0.004 mag in 
\viz, which we take to be close enough to zero that we did not choose to 
modify our overall median zero point to force them to zero.

Finally, we chose zero point corrections for $V$ and $I$ for each run
according to these offsets.  The difference of the corrections was
set to the \viz\ offset from the comparison, and the sum of the
corrections was the sum of the $V$ and $I$ offsets.  We therefore
believe that (a) our photometry is very close to Landolt and photoelectric
in zero point, (b) the error in the $V$ or $I$ photometry for
a given observation is 0.02 mag, and (c) the error in a given \viz\
color measurement is 0.017 mag (where we have divided by $\sqrt2$ to
get the error for single measurements).  We also add in quadrature 0.25 of
the zero point offset which was applied.  The offsets $\Delta V$ and 
$\Delta I$ for each run are listed in Table 1.

The reductions of \mi\ are described elsewhere (e.g., Jacoby et al.
1992).  Briefly, we fit a galaxy model to the summed, cleaned, 
sky-subtracted galaxy image and subtract it.  If there is dust present
(all too common in ellipticals and S0s as well as the large bulge
spiral galaxies we observe), we mask it out as well.  Experiments with
masking different portions of M31 and M81 (where we used $B$ band 
observations to show us clearly the location of the dust) indicate that
reasonable care in excising dust will produce a reliable \mi, both
because the extinction is less in the $I$ band and also because 
the dust tends to cause structure at relatively large scales which
are avoided by our fit to the Fourier power spectrum.  We run DoPhot
(Schechter et al. 1993)
on the resulting image to find stars, globular clusters, and background
galaxies; fit a luminosity function to the results; and derive a mask
of objects brighter than a completeness limit and an estimate of
residual variance from sources fainter than the limit.  Applying the
mask to the model-subtracted image, we calculate the variance from the
fluctuations in a number of different regions.  Finally, this variance
is converted to a value for \mi\ by dividing by the mean galaxy flux
and subtraction of the residual variance estimate from unexcised point
sources.  Generally speaking, the various estimates of \mi\ are quite
consistent from region to region, and a weighted average and error
estimate are tabulated for each observation.  If the observation was
photometric, we also record the \viz\ color found in the same region 
in which we measure \mi.

There are many galaxies for which \viz\ and \mi\ have been measured
more than once, and intercomparison of the different observations 
can be used to
evaluate whether our error estimates are reasonable.  If we 
consider all pairs and divide their difference by the 
expected error, the distribution should be a Gaussian of unity variance.
Figure 2 illustrates these distributions for \mi\ and \viz.
Evidently, the error estimates are usually quite good, with discordant
observations occurring rarely.  In most cases of discordances,
it is clear which of the observations is trustworthy, and
we simply remove the other observation from further consideration.
These excised observations occur 1.5\% of the time for \mi\ and 
0.3\% of the time for \viz, and are an indication of how frequently
bad observations occur.

After observations are averaged together, they are subjected to some final
corrections.  The mean \vi\ color of the
fluctuations is the mean of a galaxy's color \vi\ and the
``fluctuation color'' \mv--\mi, or $\vi\approx 1.85$ (since
the rms fluctuation is the square root of the flux from the galaxy
and the flux from \mi).  The value of
\mi\ is corrected according to this mean color and the color 
term for the run's photometry.  The values of \mi\ and \vi\ are
corrected for galactic extinction according to
\begin{equation}
A_V:A_{I_{\rm KC}}:E(B-V) = 3.04:1.88:1.00,
\end{equation}
where $E(B-V)$ comes from Burstein \& Heiles (1984), who give 
$A_B = 4.0\, E(B-V)$,
the relative extinction ratio $A_{I_{\rm KC}} / A_V = 0.62$ is
taken from Cohen et al. (1981) for a star halfway between an A0 and an
M star, 
and $A_V/E(B-V)$ is an adjustment of a value of 3.1 for
A0 stars common in the literature ({\it e.g.,} Cardelli, et al. 1989)
to a value of 3.04 more appropriate for early-type galaxies, following
the ratios given in Cohen et al.

The final modification is the application of K-corrections which brighten
magnitudes in $V$ and $I$ by 1.9 and $1.0 \times z$ respectively
(Schneider 1996), and brighten fluctuation magnitudes in $I$ by $7.0
\times z$ (Worthey 1996).  Note that the very red color of SBF causes
flux to be shifted rapidly out of the $I$ band with redshift, but the
\mi\ K-correction amounts to only 0.05 magnitude at a typical distance
of 2000~km/s.

\section{Calibrating \Mi}

The next step we take in trying to establish how \Mi\ varies according
to stellar population is to look at how \mi\ varies from galaxy to
galaxy within groups, where the distance to the galaxies is essentially
constant.  We originally chose to observe SBF in the $I$ band because
stellar population models indicate that \Mi\ is relatively constant
from population to population, and that the effects of age,
metallicity, and IMF are almost degenerate --- in other words, \Mi\ is 
nearly a one parameter family.  

Guided by theoretical models we seek to establish whether three 
statements are a fair description of our data:

\itemitem{(\S3.1)} \Mi\ is a one-parameter family, with a 
universal dependence on \viz\break 
(i.e., \Mi\ is a function of \viz\ with small residual scatter).

\itemitem{(\S3.2)} The zero point of the \Mi--\viz\ relation is universal.

\itemitem{(\S3.3)} The \Mi--\viz\ relation is consistent with theoretical
models of stellar populations.

To this end we chose approximately 40 nearby groups where we currently have
(or will have) observed more than one galaxy.  The groups are defined
by position on the sky and a redshift range and in most cases
correspond to one of the groups described by Faber et al. (1989).
Table 3 lists our groups.
Note that we are not trying to include all
groups, nor do we have to be complete in including all galaxies which
are members.  We are simply trying to create samples of galaxies for
which we are reasonably confident that all galaxies are at the same
distance.

\subsection{Universality of the \mi\ dependence on \viz}

Figure 3 illustrates the \mi--\viz\ relationship in six groups where we
have measured SBF in a number of galaxies: NGC~1023, Leo, Ursa Major,
Coma~I\&II, Virgo, and Fornax.  The lines are drawn with slope 4.5 and
zero point according to the fit to the data described below.  We
see that 
galaxies which meet the group criteria of position on the sky and
redshift are consistent with the same \mi--\viz\ relationship, where
the scatter reflects both the measurement error and the group depth
inferred from spread across the sky.  In Virgo we find NGC~4600 much
brighter than the rest of the galaxies, NGC~4365 significantly
fainter, and NGC~4660 (the point at $\viz=1.21$ and $\mi = 28.9$) also
with an unusually bright \mi\ for its color.  These three galaxies,
marked as smaller, square symbols, are discussed below.

Note that ellipticals and S0 galaxies are intermixed with spirals
(NGC~3368 in Leo, NGC~4548 in Virgo, NGC~891 in the NGC~1023 group,
and NGC~4565 and NGC~4725 in the Coma~I\&II group).  The two galaxies 
in Fornax marked as ``spiral'' (NGC~1373 and NGC~1386) might better
be classified as S0 on our deep CCD images.  For this admittedly small
sample we see no offset between SBF measurements in spiral bulges and
early-type galaxies.  We regard this as confirmation of our 
assumption that SBF measurements are equally valid in spiral bulges
as in early-type galaxies.

In order to test the hypothesis that \Mi\ has a universal dependence on
\viz\ in a more systematic way than fitting individual groups, we
simultaneously fit all our galaxies which match the group criteria with
\begin{equation}
\overline m_I = \langle\overline m_I^0\rangle_j + \beta \; [\viz-1.15],
\end{equation}
where we fit for values of $\langle\overline m_I^0\rangle_j$ for
each of j=1,N groups and a single value for $\beta$.  The quantity
$\langle\overline m_I^0\rangle_j$ is the group mean value for \mi\ at a
fiducial galaxy color of $\viz=1.15$.  The measurements of \viz\ and
\mi\ carry errors which the pair-wise comparisons and the averaging
procedure of section 2 indicate are accurate.

We also anticipate that there will be an irreducable ``cosmic''
scatter in \Mi.  Accordingly, in fitting \mi\ as a function of \viz, we
include an error allowance for this cosmic scatter which is nominally
0.05 magnitudes (i.e. for this fit the error in \mi\ is enhanced by
0.05 magnitude in quadrature).  In addition, we will also see scatter
because the galaxies within groups are not truly at the same distance.
We therefore calculate the rms angular position of the galaxies making
up each group, and divide this radius by $\sqrt2$ as an estimate of the
rms group depth.  Converting this to a magnitude, we add it in
quadrature to the error in \mi.  We then perform a linear fit of $N+1$
parameters which allows for errors in both the ordinate and abcissa,
according to the ``least distance'' method used by Tonry and Davis
(1981).  (This also appears in a slightly different guise in the second
edition of {\it Numerical Recipes} by Press et al. 1992)

We remove the three Virgo galaxies which we believe are at
significantly different distances from the rest of the group (NGC~4365,
NGC~4660, and NGC~4600), mindful that what is considered to be part of
Virgo and what is not is somewhat arbitrary.  We also choose to exclude
NGC~205 and NGC~5253 from the fit because recent starbursts make them
extremely blue --- we do not believe our modeling extends to such
young populations.

With 149 galaxies we have 117 degrees of freedom, and we find that
$\chi^2 = 129$, $\chi^2/N = 1.10$, and the slope of the \Mi--\viz\
relation is $4.5 \pm 0.25$.  The galaxies contributing to the fit span
a color range of $1.0<\viz<1.3$.  Because Virgo still contributes five
of the seven most discrepant points (the other two are in Cetus), the
rms depth used for Virgo ($2.35^\circ \rightarrow 0.08$~mag) may be too
small, making $\chi^2/N$ slightly bigger than one.  If we replace the 3
Virgo galaxies we omitted earlier, we find that $\chi^2/N$ rises to
1.75 for 120 degrees of freedom and the slope changes to $4.7\pm0.25$,
showing that even though these galaxies are significantly outside of
Virgo, the slope is robust.  When we experiment with adding and
removing different groups we find that the slope changes slightly, but
is always consistent with the error above.

These values for $\chi^2$ include an allowance for cosmic scatter of
0.05 magnitude and the nominal, rms group depth.  These two,
ill-constrained sources of error can play off against each other: if we
double the group depth error allowance, we get $\chi^2/N = 1.0$ for
zero cosmic scatter; if we increase the cosmic scatter to 0.10
magnitude, we need to decrease the group depth to zero in order to make
$\chi^2/N = 1.0$.  Therefore, even though we cannot unambiguously
determine how much cosmic scatter there is in the \Mi--\viz\ relation, it
appears to be $\sim$0.05 mag.

The referee pointed out that even if we make no allowance for group
depth, the cosmic scatter of 0.10 mag makes SBF the most precise
tertiary distance estimator by far, and wanted to know how sensitive
this is to our estimates of observational error.  There is not much
latitude for the cosmic scatter to be larger than 0.10 mag.  The
distribution of measurement error in \mi\ and \viz\ (which also enter
$\chi^2$) starts at 0.06 mag, and has quartiles at 0.11, 0.16, and
0.20 mag.  If we wanted to increase the cosmic scatter by $\sqrt2$ to
0.14 mag, we would have to have overestimated the observational errors
by 0.10 mag in quadrature, and apart from the fact that a quarter of
the measurements would then have imaginary errors, our pairwise
comparison of multiple observations from the previous section would
not allow such a gross reduction in observational error.

Figure 4 illustrates how \mi\ depends on \viz\ when all the group data
have been slid together by subtraction of the group mean at
$\viz=1.15$.  Note again that spiral galaxies, in this case four
galaxies with both Cepheid and SBF distances, show no offset relative
to the other early-type galaxies making up the groups in which they
appear, other than the usual trend with \viz.  The overall rms
scatter, 0.18 mag, arising from all the effects discussed above, is
a testament to the quality of SBF as a distance estimator.

The Local Group galaxies NGC~205, NGC~147, and NGC~185 have also
been plotted in Figure 4 (although they were not used in the fit), under
the assumption that they are at the same distance as M31 and M32.
This may or may not be a valid assumption for NGC~147 and NGC~185,
but they agree reasonably well with the mean relation.  In contrast to
these two galaxies, which are blue because of extremely low
metallicity, NGC~205 has undergone a recent burst of star formation
and has a strong A star spectral signature.  Because our models do
{\it not} extend to such young populations, the systematic deviation
from the mean relation is not unexpected.

The inset in Figure 4 extends the color range to show that this
deviation continues for two other galaxies where there has been recent
star formation: NGC~5253 and IC~4182.  NGC~5253 is 0.5 mag fainter than
one would expect using a naive extrapolation of the relation to its
color of $\viz = 0.84$, and IC~4182 has an SBF magnitude which is 0.75
mag fainter than one would judge from its Cepheid distance and its
color of $\viz = 0.71$.  Qualitatively this makes sense because the very
young stars change the overall color of the galaxy quite a bit but are
not very luminous in the $I$ band compared to the stars at the top of
the RGB which are the main contributors to the SBF \mi.  It may be that
these very young populations can be understood well enough that one can
safely predict the SBF absolute magnitude from the mean color, but this
is beyond the scope of this paper.

Tammann (1992) expressed concern that there are residual stellar
population effects in SBF even after the correction for \viz\ color.
However, his critique was based on an early
attempt to correlate \mi\ as a function of \viz\ (Tonry et al. 1990).
Unfortunately, that work had the wrong sign for the slope (appropriate 
for the $JHK$ bandpasses but not $I$), because it was based  on the Revised
Yale Isochrones (Green et al.  1987), which did not properly model the
line blanketing in metal rich, high luminosity stars.  The effect noted
by Tammann was a residual correlation of the corrected \mim\ with the
Mg$_2$ index among galaxies within a cluster.  Figure 5 shows these trends
do not exist for the present data and the new
\mi--\viz\ relation: in both Fornax and Virgo there is no residual
correlation with either Mg$_2$ or galaxy magnitude.

We conclude that a one-parameter, linear relation between
\Mi\ and \viz\ suffices to describe our data for $1.0 < \viz < 1.3$; the
slope of the \Mi--\viz\ relation is universally $4.5 \pm 0.25$, and we
are indeed detecting cosmic scatter in \Mi\ of order 0.05 mag.  Very few
galaxies fail to follow the relation, and for every such galaxy
at least one of the following statements is true: (1) the measurement 
of \mi\ or \viz\ is doubtful; (2) the galaxy may not be a member of the
group we assigned it to; (3) the stellar population is bluer than
$\viz=1.05$ due to recent star formation.

Note that this slope is steeper than the value of 3 tendered by Tonry
(1991) and used by Ciardullo et al. (1993) who suggested that it might
be as steep as 4.  Basically, the reason for this is that the older
data were noiser and were fitted only to errors in the ordinate,
whereas in fact the errors in \viz\ are quite significant, particularly
for the better measured \mi, which count heavily in any weighted fit.

\subsection{Universality of the \Mi\ zero point}

We have effectively tested the hypothesis that the zero point of SBF
is universal within groups, but in order to extend the test from group
to group we need independent distance estimates.  Since the groups are
all nearby, the group's redshift is not an accurate
distance estimate --- there are likely to be substantial
non-Hubble velocities included in the group's recession velocity.
We therefore turn to other distance estimators: Cepheids,
planetary nebula luminosity function (PNLF), Tully-Fisher (TF),
\dn, Type II supernovae (SNII) and Type I supernovae (SNIa).
Some of these estimators have zero points in terms of Mpc (such as
Cepheids and SNII), others have zero points in terms of km/s based on the
Hubble flow (such as \dn), and a few have both (such as TF).
For our initial discussion we seek only to establish whether the
relative distances agree with SBF; for now we do not care about the
zero point, though it will soon be addressed.

Figures 6 and 7 show the comparison between the values of the SBF
parameters \avemi\ derived previously for each of our groups and the
distances to the groups according to these 6 methods.  
The results of fitting lines of unity slope 
(allowing for errors in both coordinates) to the
data in each panel are given in Table 4.  We use the published error
estimates for all of these other indicators so $\chi^2/N$ should be
viewed with some caution: outliers and non-Gaussian errors or
over-optimistic error estimates can inflate $\chi^2/N$ even though the
mean offset is still valuable.
Since each comparison is
very important, we briefly discuss them individually.

\subsubsection{Cepheids} 

There is now a growing number of
Cepheid distances with which we compare, but we are faced with the
complication that Cepheids occur in young stellar populations, while
SBF is best measured where such populations are not present.

There are five galaxies which have
both Cepheid and SBF distances: 
NGC~224 (Freedman \& Madore 1990), 
NGC~3031 (Freedman et al 1994), 
NGC~3368 (Tanvir et al. 1995), 
NGC~5253 (Saha et al. 1995), and 
NGC~7331 (Hughes 1996).  
NGC~5253 is especially problematic for SBF, because its recent
starburst has produced a much younger and bluer stellar population than
we have calibrated.  We can, of course, also compare distances
according to group membership. There are 7 groups where this is
currently possible:
Local Group, 
M81, 
CenA, 
NGC~1023 (NGC~925 from Silbermann et al. 1996), 
NGC~3379 (also including NGC~3351 from Graham et al. 1996),
NGC~7331, and 
Virgo (including NGC~4321 from Ferrarese et al. 1996,
NGC~4536 from Saha et al. 1996a, and NGC~4496A from Saha et al. 1996b;
we exclude NGC~4639 from Sandage et al. 1996 because we are also
excluding NGC~4365 and the W cloud from the SBF mean).  In the former
case we find that fitting a line to \avemi\ as a function of
\dmod\ yields a mean offset of $-1.75\pm0.05$ mag with $\chi^2/N$ of
3.4 for 4 degrees of freedom, and $-1.82\pm0.06$ mag with $\chi^2/N$ of
0.3 for 3 degrees of freedom when NGC~5253 is excluded.  In the latter
case we get a mean offset of $-1.74\pm0.05$ mag with $\chi^2/N$ of 0.6
for 6 degrees of freedom.  When NGC~5253 is excluded, the rms scatter
is remarkably small, only 0.12 magnitudes for the galaxy comparison and
0.16 magnitudes for the group comparison.

\subsubsection{PNLF}
Ciardullo et al. (1993) reported virtually perfect agreement between
SBF and PNLF, but recent publications (Jacoby et al. 1996) have raised some
discrepancies.  Examination of Figure 7 reveals that our fit has two
outliers: Coma~I (e.g. NGC~4278) and Coma~II (e.g. NGC~4494).  Because
we do not know how to resolve this issue at present, Table 4 gives the
result for $\avemi - \dmod_{\hbox{PNLF}}$ for the entire sample and
when these two outliers are removed.  Since PNLF is fundamentally
calibrated on Cepheids, this is not independent of the previous
number, but it does confirm that PNLF and SBF are measuring the same
relative distances.

\subsubsection{SNII} The expanding photospheres method (EPM)
described most recently by 
Eastman et al. (1996) offers distance estimates which are largely
independent of the Cepheid distance scale.  There is only one galaxy
with both an EPM and an SBF distance (NGC~7331), but there have also been
two SNII in Dorado (NGC~1559 and NGC~2082), two in Virgo (NGC~4321 and
NGC~4579), and one in the NGC~1023 group (NGC~1058).  The agreement
between EPM and SBF (Fig. 6) is good.  The farthest outlier is
NGC~7331, for which SBF and Cepheid distances are discordant with the
SNII distance. Table 4 lists separately the zero point, scatter, and
$\chi^2/N$ when NGC~7331 is included and excluded.

\subsubsection{TF (Mpc calibration)} B. Tully (1996) was kind
enough to provide us with TF distances to the SBF groups in advance of
publication.  The fit between TF and SBF gives $\avemi - \dmod =
-1.69\pm0.03$ mag.  This is again not independent of the Cepheid
number, since the TF zero point comes from the same Cepheid
distances.   Figure 7 demonstrates that the agreement is
generally good, despite the high $\chi^2/N$ which comes from a few
non-Gaussian outliers.  We cannot tell whether these outliers reflect
non-Gaussian errors in the methods or simply the difficulties of
choosing spirals and early type galaxies in the same groups.

\subsubsection{TF (km/s calibration)} We applied the
SBF group criteria to the ``Mark II'' catalog of galaxy distances
distributed by D. Burstein.  We selected all galaxies with ``good'' TF
distances (mostly from Aaronson et al.  1982) and computed an average
distance to the groups, applying the usual Malmquist bias correction
according to the precepts of Lynden-Bell et al. (1988) and the
error estimates from Burstein.  Because these
distances have a zero point based on the distant Hubble flow, we derive
an average offset of $\avemi\ - 5\log d = 13.55\pm0.08$ mag.

\subsubsection{\dn} Most of the SBF groups are the same as
those defined by Faber et al.  (1989).  We compare their Malmquist bias
corrected distances to these groups (which are based on a zero point
from the distant Hubble flow) with SBF and find the same result as
Jacoby et al. (1992): the distribution of errors has a larger tail than
Gaussian, but the error estimates accurately describe the central core
of the distribution.  $\chi^2/N$ is distinctly larger than 1, but the
difference histogram in Figure 7 reveals that this is because of the
tails of the distribution.  The fit between \dn\ and SBF gives
$\avemi\ - 5\log d = 13.64\pm0.05$ mag.

\subsubsection{SNIa}

Extraordinary claims have been made recently about the quality of SNIa
as distance estimators.  Some authors (e.g. Sandage and Tammann 1993)
claim that suitably selected (``Branch normal'') SNIa are standard
candles with a dispersion as little as 0.2 mag.  Others (e.g.  Phillips
1993) believe that they see a correlation between SNIa luminosity and
their rate of decline, parametrized by the amount of dimming 15
days after maximum, $\Delta m_{15}$.  Still others (e.g. Riess et al.
1995) agree with Phillips (1993) but believe that they can categorize
SNIa better by using more information about the light curve shape than
just this rate of decline.  Finally, there is the ``nebular SNIa
method'' of Ruiz-Lapuente (1996) which tries to determine the mass of
the exploding white dwarf by consideration of the emission lines from
the expanding ejecta.  We therefore choose to compare SBF
distances with SNIa under two assumptions: that SNIa are standard
candles, and that $m_{max} - \alpha\Delta m_{15}$ is a better indicator
of distance.  In both cases we restrict our fits to $0.8 < \Delta
m_{15} < 1.5$ as suggested by Hamuy et al. (1995) and use a distance
error of 0.225 mag for each SNIa.

SNIa have been carefully tied to a zero point according to the distant
Hubble flow (one of the main advantages of SNIa) by Hamuy et al.
(1995), under both assumptions.  There have also been vigorous attempts
to tie the SNIa to the Cepheid distance scale which we have chosen not to
use because of the circularity with our direct comparison between SBF
and Cepheids.

The results are both encouraging and discouraging.  We find that there
is indeed a good correlation between SNIa distance and SBF, with
average values of $\avemi\ - 5\log d = 13.92\pm0.08$ mag and
$14.01\pm0.08$ mag for the group comparison under the two assumptions.
As illustrated in Figure 6, $m_{max} - \alpha\Delta m_{15}$ does
correlate better with distance than $m_{max}$, but as long as ``fast
declining'' SNIa are left out there is scant difference between the
zero point according to the two methods.

The panels of Figure 6 showing SBF and SNIa hint at a systematic
change between the nearest three and the farthest three groups, in the
sense that there appears to be a change in zero point by about 0.7 mag.
One might worry that this is evidence that SBF is ``bottoming out'',
but there is no hint of this in the comparisons with TF and \dn\ in
Figure 7 which extend to much fainter \mi.  One might also worry about
whether there are systematic differences in SNIa in spirals and
ellipticals, and biases from the lack of nearby ellipticals or S0
galaxies.  However, it is probably premature to examine these points
in too much detail.  For example, the point at $\avemi \approx 28$ uses
the SBF distance to Leo~I, but the SNIa occurred in NGC~3627 which
lies $8^\circ$ away from the Leo~I group.  This is a fundamental
difficulty in the SBF--SNIa comparison, which will improve as SBF
extends to greater distances and more nearby SNIa are observed.

There are seven galaxies bearing SNIa where SBF distances have been
measured: NGC~5253 (SN~1972E), NGC~5128 (SN~1986G), NGC~4526
(SN~1994D), NCG~2962 (SN~1995D), NGC~1380 (SN~1992A), NGC~4374
(SN~1991bg), NGC~1316 (SN~1980N).  Inasmuch as two of these are slow
decliners (SN~1986G, SN~1991bg), we fit the remaining five using the
SBF distance to the galaxy instead of the group. We derive $\avemi\ -
5\log d = 13.86\pm0.12$ mag and $14.01\pm0.12$ mag for the two
methods.

We regard the SBF distance to NGC~5253 as uncertain because we
have not calibrated \Mi\ for such a young stellar population.  We
thus also recompare SBF and SNIa with NGC~5253 removed from
consideration.  $\chi^2/N$ becomes dramatically smaller in both cases
and $\avemi\ - 5\log d$ become smaller by about 0.2 mag to
$13.64\pm0.13$ mag and $13.87\pm0.13$ mag.

\subsubsection{Zero point summary}

These comparisons demonstrate that the second hypothesis is
correct: the zero point of the \Mi--\viz\ relationship is
universal.  We use the SBF--Cepheid fit to derive a final,
empirical relationship between \mi\ and \viz:
\begin{equation}
\overline M_I = (-1.74\pm0.07)\, +\, (4.5\pm0.25)\, [\viz - 1.15].
\end{equation}

This zero point differs from that of Tonry (1991) by about 0.35
magnitude.  The reason is simply that the 1991 zero point was based
entirely on M31 and M32, and the observational error in both \mi\ and
\viz\ worked in the same direction, as did the photometric zero point
errors (cf. Table 1 for K0990).  The SBF distances which have been
published therefore increase by about 15 percent (for example Fornax
moves from 15~Mpc to 17~Mpc), except for Virgo,
where the earlier result included NGC~4365 which we now exclude
in calculating the average distance to the core of the cluster.  This
new calibration is based on 10 Cepheid distances in 7 groups and 44
SBF distances.  As seen in Figure 6 and Table 4, these are highly
consistent with one another with a scatter of about 0.15 mag.
Along with the extensive photometric recalibration, this zero point
should be accurate to $\pm0.07$ mag.  This error estimate
makes an allowance of 0.05 mag for the uncertainty in the Cepheid zero
point in addition to the statistical error of 0.05 mag, and
the comparisons with theory and SNII give us confidence that this
truly is correct.

\subsection{Comparison with theory}

Finally we test our third hypothesis by comparing our \Mi--\viz\
relationship with theoretical models of stellar populations.
Figure 8 shows the model predictions of Worthey (1993a,b) along
with the empirical line.  
When the theoretical models are fitted with the empirically determined
slope of 4.5, they yield a theoretical zero point of $-1.81$ mag with
an rms scatter of $0.11$ mag for the SBF relation.  We enter this
value in Table 4, with the scatter offered as an ``error estimate'',
but it must be remembered that this is fundamentally different from
the other entries in the table.
There is good agreement here, although
the theoretical result for \Mi\ may be slightly brighter (0.07 mag) or
slightly redder (0.015 mag) than the empirical result.  Given the
difficulties that the theoretical models have in simultaneously
fitting the color and Mg$_2$ indices of real galaxies, we regard this
agreement as excellent confirmation of the empirical calibration.

\section{The Hubble Constant}

The scope of this paper does not extend to comparing SBF distances
with velocity; this will be the subject of the next paper in the
series.  However, the comparison with other distance estimators 
does provide us with a measurement of the Hubble constant.

The comparison with other estimators whose zero point is defined in
terms of Mpc tells us the absolute magnitude of SBF.  At our fiducial
color of $\viz=1.15$, we find that Cepheids give us an absolute magnitude
$\Mi = -1.74\pm0.05$.  We prefer the group-based Cepheid comparison
because of the very few SBF measurements possible in spirals which
have Cepheids.
The other Mpc-based distance estimators are all
consistent with this zero point, as we would hope since they are
calibrated with the same Cepheid data.
The results from theoretical models of stellar populations and SNII are
also consistent with this zero point, and provide independent
confirmation of the validity of the Cepheid distance scale.

The comparison of SBF with estimators whose zero point is based
on the large scale Hubble flow is less consistent.  The estimators
based on galaxy properties, TF and \dn, are consistent with
one another and consistent with SBF in terms of relative distances.
They give a zero point for SBF at the fiducial color of $\viz=1.15$ of
$\avemi = 5\log d\hbox{(km/s)} + 13.59\pm0.07$, where the error comes
from the rms divided by $\sqrt{N-1}$.

Supernovae and SBF are more interesting.  The group membership of the
Cepheid galaxies was not difficult since they were specifically chosen
to be group members.  In contrast, the SNIa are not easy to assign to
groups in many cases.  Depending on (1) whether we fit galaxies
individually or groups, (2) whether we use the
``standard candle'' model for SNIa or the ``light curve decline''
relation, and (3) whether we include or exclude NGC~5253 for which we
regard our stellar population calibration as unknown, we get values
for the SBF zero point as low as 13.64 and as high as 14.01 (Table 4).
Averaging the two methods and again estimating uncertainties from rms
divided by $\sqrt{N-1}$, we find $13.96\pm0.17$ for groups and
$13.75\pm0.14$ for galaxies.  Because these differ from the TF and
\dn\ by $2.0\;\sigma$ and $1.0\;\sigma$ respectively, the discrepancy
may not be statistically significant.

It is possible that there are systematic errors in the tie to the
distant Hubble flow for TF and \dn, whereas the SNIa appear to be
wonderfully consistent with the large scale Hubble flow.  On the other
hand, the nearby SNIa do not agree with SBF or Cepheids as well as one
might hope from the scatter against the Hubble flow, which makes one
worry about the systematics with SN1a.  For example, the SN1a
distances predicted for the Fornax clusters are significantly larger
than the very recent Cepheid measurement of the distance to the Fornax
cluster (Silbermann et al. 1996b).  SNII appear to agree pretty well
with SBF and Cepheids, and there should eventually be enough of them
to tie very well to the large scale Hubble flow.  In subsequent papers
we will present the direct tie between SBF and the Hubble flow, both
from ground-based observations as well as HST observations beyond
5000~km/s, but at present we depend on these other estimators to
tie to the Hubble flow.  It is therefore
with some trepidation that we offer a value for $H_0$.

We have a calibration for \Mi; we have several calibrations for \mi\
in terms of $5\log d\hbox{(km/s)}$; and of course $\dmod = 5\log
d\hbox{(km/s)} + 25 - 5\log H_0$.  If we use the TF and \dn\
calibration of SBF we get $H_0 = 86$~\hbox{km/s/Mpc}.  Examining groups
and averaging the ``standard candle'' and the ``$\Delta m_{15}$''
assumptions about SNIa gives us $H_0 = 72$~\hbox{km/s/Mpc}.  If we
compare galaxies directly without resorting to group membership, but
leave out NGC~5253, we get an average $H_0 = 80$~\hbox{km/s/Mpc}.

We suspect that there is more to the SNIa story than is currently
understood, so we therefore prefer not to use it to the exclusion of
all other distance estimators.  The range we find for $H_0$ is 
$$ H_0 = 72 - 86\;\hbox{km/s/Mpc,}$$
and our best guess at this point is derived by averaging the ties
to the Hubble flow from TF, \dn, SNIa (both methods) in groups
and SNIa (both methods) galaxy by galaxy.  This weights the SNIa
slightly more heavily than TF and \dn\ and gives a zero point of 13.72
which translates to
$$ H_0 = 81\pm6\;\hbox{km/s/Mpc.}$$
The final error term includes a contribution of 0.07 magnitude from the
disagreement between the Cepheid and theory zero points (which we hope
is indicative of the true accuracy of our calibrations), and an
allowance of 0.13 magnitude for the uncertainty in the tie to the
distant Hubble flow (judged from the scatter among the various
methods).

In order to facilitate comparisons with SBF distances, we offer the
SBF distance to 12 nearby groups in Table 5.  The relative distances
are completely independent of any other distance estimator, and the
zero point uses our Cepheid-based calibration.  As we finish our
reductions and analysis, the remainder of the group and individual
galaxy distances will be published.  

\section{Summary and Conclusions}

We have described the observational sample which comprises the SBF
Survey of Galaxy Distances.  The survey was conducted over numerous
observing runs spanning a period of nearly seven years.  The photometry
of the sample has been brought into internal consistency by applying
small systematic corrections to the photometric zero points of the
individual runs.  Based on comparisons between overlapping galaxy
observations, we find that our error estimates for $(V-I)$ and
$\overline m_I$ are reliable, after correction for the
photometry offsets.

From our measurements of \mi\ within galaxy groups, we 
conclude that \Mi\ is
well described by a linear function of \viz.
Comparison of our relative distances with Cepheid distances to these 
groups indicates that this linear relationship is universal and yields
the zero point calibration for the SBF method.
This calibration is applicable to galaxies that are in the color range
$1.0 < \viz < 1.3$ and which have not experienced
recent bursts of star formation.  Any intrinsic, or ``cosmic,''
scatter about this relation is small, of order 0.05~mag.
Owing to many more data and improved photometry, this
new calibration differs in its zero point by 0.35 mag from the
earlier one of Tonry (1991), but is much closer to Worthey's (1993)
theoretical zero point, differing by just 0.07~mag. 
We take this close agreement
to be an independent confirmation of the Cepheid distance scale.

An extensive set of comparisons between our SBF distances and those
estimated using other methods provides still further evidence for the
universality of the \Mi--\viz\ relation.  
We find that the various methods are all generally quite reliable,
apart from occasional outliers which serve to inflate the 
$\chi^2$ values for the comparisons.  
Coupled with our distance zero point, our comparisons with
methods tied to the distant Hubble flow yield values of $H_0$ in
the range 72--86 km/s/Mpc.  The comparison with SNIa suggests values
between 72 and 80, and 
\dn\ and TF call for values around 86.
Thus, the controversy over $H_0$ continues, but the famous ``factor
of two'' is now a factor of 20 percent.

Although the SBF Survey is still a work in progress, it is near enough to
completion that the calibration presented in this paper
should not change in any significant way.  
Future papers in this series will use the SBF survey distances to address
such issues as the velocity field of
the Local Supercluster and a direct determination of $H_0$, bulk flows,
the Great Attractor, and the specific details of our SBF analysis method,
including comprehensive listings of our \viz\ and distance measurements
for individual galaxies.

\acknowledgments

We would like to thank many people for collecting SBF and photometry
observations for us: Bob Barr, Andre Fletcher, Xiaohui Hui, Gerry
Luppino, Mark Metzger, Chris Moore, and Paul Schechter.
This research was supported by NSF grant AST-9401519,
and AD acknowledges the support of the National Science Foundation
through grant AST-9220052.

\clearpage
\begin{deluxetable}{llllllllllll}
\small
\tablecaption{Observing Runs.\label{tbl1}}
\tablewidth{0pt}
\tablehead{
\colhead{Run} & \colhead{Telescope}   & \colhead{CCD}   & 
\colhead{$^{\prime\prime}/p$}  & 
\colhead{$m_{1V}$}  & \colhead{$A_V$} & \colhead{$C_V$} &
\colhead{$\Delta V$}& 
\colhead{$m_{1I}$}  & \colhead{$A_I$} & \colhead{$C_I$} &
\colhead{$\Delta I$}
} 
\startdata
K0389 & KPNO4m & TI-2 & 0.299 & 26.15 & 0.150 & \llap{$-$}0.070 &  0.014 & 25.42 &  0.070 & 0.000 & 0.012 \nl
M1189 & MDM2.4m& ACIS & 0.465 & 23.47 & 0.179 &  0.013 &  0.000 & 22.44 &  0.065 & 0.000 & 0.000 \nl
C0990 & CTIO4m & TI   & 0.299 & 26.23 & 0.160 &  0.0   & \llap{$-$}0.026 & 25.29 &  0.080 & 0.0   &\llap{$-$}0.003 \nl
K0990 & KPNO4m & TI-2 & 0.299 & 26.26 & 0.160 &  0.0   &  0.019 & 25.39 &  0.080 & 0.0   & 0.045 \nl
H0291 & CFH3.6m& SAIC & 0.131 & 24.86 & 0.089 &  0.0   & \llap{$-$}0.016 & 24.62 &  0.033 & 0.0   & 0.029 \nl
L0391 & LCO2.4m& Tek  & 0.229 & 24.92 & 0.15  &  0.0   &  0.024 & 24.62 &  0.07  & 0.0   & 0.030 \nl
C0491 & CTIO4m & Tek1 & 0.472 & 26.06 & 0.16  &  0.0   &  0.069 & 26.02 &  0.11  & 0.0   & 0.061 \nl
K0691 & KPNO4m & TI-2 & 0.300 & 25.97 & 0.155 &  0.0   & \llap{$-$}0.002 & 25.36 &  0.06  & 0.0   & 0.019 \nl
C1091 & CTIO4m & Tek2 & 0.472 & 26.21 & 0.45  & \llap{$-$}0.007 &  0.019 & 26.08 &  0.3   & 0.025 & 0.040 \nl
H1091 & CFH3.6m& SAIC & 0.131 &       &       &        &  0.000 & 24.62 &  0.07  & 0.0   & 0.000 \nl
L1191 & LCO2.4m& Tek  & 0.229 & 24.92 & 0.15  &  0.0   &  0.034 & 24.62 &  0.07  & 0.0   &\llap{$-$}0.014 \nl
M1191 & MDM2.4m& ACIS & 0.257 & 25.11 & 0.205 &  0.0   & \llap{$-$}0.025 & 24.53 &  0.102 & 0.035 & 0.007 \nl
C0492 & CTIO4m & Tek2 & 0.472 & 26.05 & 0.220 &  0.005 &  0.014 & 25.93 &  0.145 & 0.030 &\llap{$-$}0.020 \nl
M0492 & MDM2.4m& Lorl & 0.343 & 24.69 & 0.33  &  0.000 &  0.010 & 24.84 &  0.20  & 0.045 & 0.007 \nl
H0592 & CFH3.6m& STIS & 0.152 & 25.91 & 0.210 &  0.0   & \llap{$-$}0.010 & 25.60 &  0.110 & 0.0   &\llap{$-$}0.038 \nl
M0892 & MDM2.4m& Lorl & 0.343 & 24.64 & 0.254 &  0.000 &  0.000 & 24.74 &  0.145 & 0.045 & 0.000 \nl
L1092 & LCO2.4m& Tek  & 0.229 & 24.92 & 0.15  &  0.0   &  0.000 & 24.62 &  0.07  & 0.0   & 0.000 \nl
M1092 & MDM2.4m& Lorl & 0.343 & 24.68 & 0.32  &  0.000 &  0.010 & 24.82 &  0.22  & 0.046 & 0.029 \nl
L0493 & LCO2.4m& Tek  & 0.229 & 24.92 & 0.15  &  0.0   &  0.027 & 24.62 &  0.07  & 0.0   &\llap{$-$}0.070 \nl
M0493 & MDM2.4m& Lorl & 0.343 & 24.67 & 0.21  &  0.022 & \llap{$-$}0.004 & 24.35 &  0.13  & 0.030 & 0.014 \nl
M0493 & MDM2.4m& Tek  & 0.275 & 25.32 & 0.24  &  0.005 & \llap{$-$}0.004 & 24.60 &  0.11  & 0.012 & 0.014 \nl
M0593 & MDM2.4m& Lorl & 0.343 & 24.70 & 0.22  &  0.022 & \llap{$-$}0.033 & 24.35 &  0.138 & 0.030 &\llap{$-$}0.009 \nl
M0593 & MDM2.4m& Tek  & 0.275 & 25.32 & 0.198 &  0.025 & \llap{$-$}0.033 & 24.74 &  0.134 & 0.030 &\llap{$-$}0.009 \nl
M0893 & MDM2.4m& Lorl & 0.343 & 24.82 & 0.19  &  0.012 & \llap{$-$}0.021 & 24.54 &  0.10  & 0.025 &\llap{$-$}0.006 \nl
M0294 & MDM2.4m& Tek  & 0.275 & 25.11 & 0.150 & \llap{$-$}0.017 & \llap{$-$}0.040 & 24.84 &  0.058 & 0.015 &\llap{$-$}0.074 \nl
L0394 & LCO2.4m& Tek  & 0.229 & 24.92 & 0.15  &  0.0   &  0.026 & 24.62 &  0.07  & 0.0   &\llap{$-$}0.030 \nl
L0994 & LCO2.4m& Tek  & 0.229 & 24.92 & 0.15  &  0.0   &  0.016 & 24.62 &  0.07  & 0.0   & 0.022 \nl
G0395 & MDM1.3m& Lorl & 0.637 & 23.22 & 0.15  &  0.019 & \llap{$-$}0.029 & 22.84 &  0.064 & 0.026 &\llap{$-$}0.007 \nl
G0495 & MDM1.3m& Lorl & 0.637 & 23.15 & 0.15  &  0.015 & \llap{$-$}0.041 & 22.83 &  0.064 & 0.010 &\llap{$-$}0.012 \nl
G0495 & MDM1.3m& STIS & 0.445 & 23.11 & 0.15  &  0.015 & \llap{$-$}0.041 & 22.65 &  0.100 & 0.010 &\llap{$-$}0.012 \nl
G0695 & MDM1.3m& Lorl & 0.637 & 23.13 & 0.15  &  0.042 & \llap{$-$}0.015 & 22.79 &  0.061 & 0.026 & 0.000 \nl
G0995 & MDM1.3m& Lorl & 0.637 & 22.97 & 0.14  &  0.014 & \llap{$-$}0.009 & 22.81 &  0.05  & 0.026 &\llap{$-$}0.015 \nl
G1095 & MDM1.3m& STIS & 0.445 & 22.98 & 0.15  &  0.005 & \llap{$-$}0.015 & 22.70 &  0.06  & 0.008 &\llap{$-$}0.015 \nl
M0295 & MDM2.4m& Tek  & 0.275 & 25.11 & 0.150 & \llap{$-$}0.014 & \llap{$-$}0.040 & 24.84 &  0.058 & 0.012 &\llap{$-$}0.074 \nl
M0395 & MDM2.4m& Tek  & 0.275 & 25.11 & 0.150 & \llap{$-$}0.017 & \llap{$-$}0.040 & 24.84 &  0.058 & 0.015 &\llap{$-$}0.074 \nl
L0495 & LCO2.4m& Tek  & 0.229 & 24.92 & 0.15  &  0.0   &  0.008 & 24.62 &  0.07  & 0.0   &\llap{$-$}0.027 \nl
L1095 & LCO2.4m& Tek  & 0.229 & 24.92 & 0.15  &  0.0   &  0.000 & 24.62 &  0.07  & 0.0   & 0.000 \nl
M1295 & MDM2.4m& Tek  & 0.275 & 25.11 & 0.150 & \llap{$-$}0.017 & \llap{$-$}0.040 & 24.84 &  0.058 & 0.015 &\llap{$-$}0.074 \nl
M0196 & MDM2.4m& Tek  & 0.275 & 25.11 & 0.150 & \llap{$-$}0.017 & \llap{$-$}0.040 & 24.84 &  0.058 & 0.015 &\llap{$-$}0.074 \nl
M0396 & MDM2.4m& Tek  & 0.275 & 25.11 & 0.150 & \llap{$-$}0.017 & \llap{$-$}0.040 & 24.84 &  0.058 & 0.015 &\llap{$-$}0.074 \nl
\enddata
\tablecomments{Columns: 
Run name, telescope, detector, plate scale (\arcsec/pixel),
photometric zero point, extinction, color term, and run offset for the
$V$ band then $I$ band.}

\end{deluxetable}

\begin{deluxetable}{lrrrrrr}
\tablecaption{Landolt Fields.\label{tbl2}}
\tablewidth{0pt}
\tablehead{
\colhead{Field} &  \colhead{RA} & \colhead{Dec}  &
\colhead{$V_{min}$}  & \colhead{$V_{max}$} & 
\colhead{$\vi_{min}$} & \colhead{$\vi_{max}$}
} 
\startdata
SA92-250    & 00 54 41 & $+$00 41 11 & 14.09 & 15.35 &    0.67 & 1.34 \nl
SA95-190    & 03 53 16 & $+$00 16 25 & 12.63 & 14.34 &    0.42 & 1.37 \nl
SA95-275    & 03 54 40 & $+$00 27 24 & 12.17 & 14.12 &    1.40 & 2.27 \nl
SA98-650    & 06 52 11 & $-$00 19 23 & 11.93 & 13.75 &    0.17 & 2.09 \nl
Rubin-149   & 07 24 13 & $-$00 31 58 & 11.48 & 13.87 & $-$0.11 & 1.13 \nl
PG0918+029  & 09 21 36 & $+$02 47 03 & 12.27 & 14.49 & $-$0.29 & 1.11 \nl
PG1323$-$085  & 13 25 44 & $-$08 49 16 & 12.08 & 14.00 & $-$0.13 & 0.83 \nl
PG1633+099  & 16 35 29 & $+$09 46 54 & 12.97 & 15.27 & $-$0.21 & 1.14 \nl
SA110-232   & 18 40 50 & $+$00 01 51 & 12.52 & 14.28 &    0.89 & 2.36 \nl
SA110-503   & 18 43 05 & $+$00 29 10 & 11.31 & 14.20 &    0.65 & 2.63 \nl
Markarian-A & 20 43 59 & $-$10 47 42 & 13.26 & 14.82 & $-$0.24 & 1.10 \nl
\enddata
\tablecomments{Columns: 
Field name, J2000 coordinates, $V$ magnitude of the brightest and
faintest star, and the \vi\ colors of the bluest
and reddest star.}

\end{deluxetable}

\begin{deluxetable}{llrrrrrrr}
\tablecaption{Nearby SBF Groups.\label{tbl3}}
\tablewidth{0pt}
\tablehead{
\colhead{Group} &  \colhead{Example}  &
\colhead{RA}  & \colhead{Dec} & \colhead{rad} & 
\colhead{$v_{ave}$} & \colhead{$v_{min}$}& \colhead{$v_{max}$} &
\colhead{7S\#}
} 
\startdata
LocalGroup & N0224 &  10.0 &  41.0 &   5 & \llap{$-$}300 & \llap{$-$}500 & \llap{$-$}100 & 282 \nl
Cetus      & N0636 &  24.2 &  \llap{$-$}7.8 &  10 & 1800 & 1500 & 2000 &  26 \nl
N1023      & N1023 &  37.0 &  35.0 &   9 &  650 &  500 & 1000 &     \nl
N1199      & N1199 &  45.3 & \llap{$-$}15.8 &   2 & 2700 & 2500 & 3000 &  29 \nl
Eridanus   & N1407 &  53.0 & \llap{$-$}21.0 &   6 & 1700 &  500 & 2300 &  32 \nl
Fornax     & N1399 &  54.1 & \llap{$-$}35.6 &   6 & 1400 &  500 & 2100 &  31 \nl
Dorado     & N1549 &  63.7 & \llap{$-$}55.7 &   5 & 1300 &  700 & 1700 & 211 \nl
N1700      & N1700 &  72.2 &  \llap{$-$}3.5 &   3 & 4230 & 3600 & 4500 & 100 \nl
N2768      & N2768 & 136.9 &  60.2 &   4 & 1360 & 1100 & 1700 & 215 \nl
M81        & N3031 & 147.9 &  69.3 &   8 &  \llap{$-$}40 & \llap{$-$}200 &  400 & \nl
N3115      & N3115 & 150.7 &  \llap{$-$}7.5 &   8 &  700 &  100 &  900 & \nl
LeoIII     & N3193 & 153.9 &  22.1 &   3 & 1400 & 1000 & 1700 &  45 \nl
LeoI       & N3379 & 161.3 &  12.8 &   2 &  900 &  500 & 1200 &  57 \nl
LeoII      & N3607 & 168.6 &  18.3 &   3 &  950 &  650 & 1500 &  48 \nl
N3640      & N3640 & 169.6 &   3.5 &   2 & 1300 & 1200 & 1800 &  50 \nl
UMa        & N3928 & 180.0 &  47.0 &   8 &  900 &  700 & 1100 & \nl
N4125      & N4125 & 181.4 &  65.5 &   3 & 1300 & 1000 & 1700 &  54 \nl
VirgoW     & N4261 & 184.2 &   6.1 &   2 & 2200 & 2000 & 2800 & \nl
ComaI      & N4278 & 184.4 &  29.6 &   3 & 1000 &  200 & 1400 &  55 \nl
CVn        & N4258 & 185.0 &  44.0 &   7 &  500 &  400 &  600 & \nl
N4386      & N4386 & 185.6 &  75.8 &   5 & 1650 & 1500 & 2100 &  98 \nl
N4373      & N4373 & 185.7 & \llap{$-$}39.5 &   2 & 3400 & 2500 & 3800 &  35 \nl
Virgo      & N4486 & 187.1 &  12.7 &  10 & 1150 & \llap{$-$}300 & 2000 &  56 \nl
ComaII     & N4494 & 187.2 &  26.1 &   5 & 1350 & 1200 & 1400 & 235 \nl
N4594      & N4594 & 189.4 & \llap{$-$}11.4 &   5 & 1100 &  900 & 1200 & \nl
M51        & N5194 & 200.0 &  45.0 &   4 &  480 &  380 &  580 & \nl
Centaurus  & N4696 & 191.5 & \llap{$-$}41.0 &   3 & 3000 & 2000 & 5000 &  58 \nl
CenA       & N5128 & 200.0 & \llap{$-$}39.0 &  15 &  550 &  200 &  600 & 226 \nl
N5322      & N5322 & 212.5 &  57.0 &   6 & 2000 & 1600 & 2400 & 245 \nl
N5638      & N5638 & 216.0 &   3.5 &   3 & 1650 & 1400 & 1900 &  68 \nl
N5846      & N5846 & 226.0 &   1.8 &   2 & 1700 & 1200 & 2200 &  70 \nl
N5898      & N5898 & 228.8 & \llap{$-$}23.9 &   2 & 2100 & 2000 & 2700 &  71 \nl
N6684      & N6684 & 281.0 & \llap{$-$}65.2 &  10 &  850 &  500 & 1200 &  78 \nl
N7144      & N7144 & 327.4 & \llap{$-$}48.5 &   6 & 1900 & 1500 & 2000 &  84 \nl
N7180      & N7180 & 329.9 & \llap{$-$}20.8 &  10 & 1500 & 1300 & 1900 & 265 \nl
N7331      & N7457 & 338.7 &  34.2 &   9 &  800 &  800 & 1100 & \nl
Grus       & I1459 & 343.6 & \llap{$-$}36.7 &   5 & 1600 & 1400 & 2300 & 231 \nl 
\enddata
\tablecomments{Columns: 
Group name, sample member, RA and Dec (B1950), group radius (deg), 
mean heliocentric velocity, minimum and maximum velocities for
inclusion in the group, and  group number from Faber et al. (1989)}

\end{deluxetable}

\begin{deluxetable}{lccrrrrrl}
\tablecaption{Distance Comparisons.\label{tbl4}}
\tablewidth{0pt}
\tablehead{
\colhead{Estimator} &  \colhead{Grp/gxy}  &
\colhead{Distance}  & \colhead{N} & 
\colhead{$\langle\overline{m}_I^0\rangle-d$} & \colhead{$\pm$} &
\colhead{rms} & \colhead{$\chi^2/N$} & \colhead{Comments}
} 
\startdata
Cepheid    &  Grp &  (m-M) &  7 & \llap{$-$}1.74 & 0.05 & 0.16 & 0.6 & \nl
Cepheid    &  gxy &  (m-M) &  5 & \llap{$-$}1.75 & 0.06 & 0.33 & 3.4 & \nl
Cepheid    &  gxy &  (m-M) &  4 & \llap{$-$}1.82 & 0.07 & 0.12 & 0.3 & less N5253 \nl
PNLF       &  Grp &  (m-M) & 12 & \llap{$-$}1.63 & 0.02 & 0.33 & 7.5 & \nl
PNLF       &  Grp &  (m-M) & 10 & \llap{$-$}1.69 & 0.03 & 0.20 & 2.2 & less ComaI/II \nl
SNII       &  Grp &  (m-M) &  5 & \llap{$-$}1.80 & 0.12 & 0.36 & 1.4 & \nl
SNII       &  Grp &  (m-M) &  4 & \llap{$-$}1.76 & 0.12 & 0.22 & 1.1 & less N7331 \nl
TF         &  Grp &  (m-M) & 26 & \llap{$-$}1.69 & 0.03 & 0.41 & 2.1 & \nl
TF (MkII)  &  Grp &  5logd & 29 & 13.55 & 0.08 & 0.59 & 2.1 & \nl
Dn-sigma   &  Grp &  5logd & 28 & 13.64 & 0.05 & 0.44 & 1.9 & \nl
SNIa ($M_{max}$)&  Grp &  5logd &  6 & 13.92 & 0.08 & 0.38 & 3.6 & \nl
SNIa ($\Delta m_{15}$)&  Grp &  5logd &  6 & 14.01 & 0.08 & 0.40 & 3.6 & \nl
SNIa ($M_{max}$)&  gxy &  5logd &  5 & 13.86 & 0.12 & 0.54 & 4.9 & \nl
SNIa ($\Delta m_{15}$)&  gxy &  5logd &  5 & 14.01 & 0.12 & 0.43 & 3.2 & \nl
SNIa ($M_{max}$)&  gxy &  5logd &  4 & 13.64 & 0.13 & 0.22 & 1.0 & less N5253 \nl
SNIa ($\Delta m_{15}$)&  gxy &  5logd &  4 & 13.87 & 0.13 & 0.30 & 1.8 & less N5253 \nl
Theory     &      &        &    & \llap{$-$}1.81 &      & 0.11 &     & \nl
\enddata
\tablecomments{Columns: 
Name of the estimator, comparison
by group or by galaxy, estimator's zero point based on Mpc \dmod\ or
Hubble flow (5logd~km/s), number of comparison points, 
mean difference between SBF and the estimator,
expected error in this mean based on error estimates,
rms scatter in the comparison, $\chi^2/N$, and comments.}

\end{deluxetable}

\begin{deluxetable}{llrrrrrrrl}
\tablecaption{SBF Distances to Groups.\label{tbl5}}
\tablewidth{0pt}
\tablehead{
\colhead{Group} &  \colhead{Example}  &
\colhead{RA}  & \colhead{Dec} & 
\colhead{$v_{ave}$} & \colhead{$N$} & \colhead{\dmod} &
\colhead{$\pm$} & \colhead{$d$} & $\pm$
} 
\startdata
LocalGrp & N0224 &  10.0 &  41.0 &  \llap{$-$}300 &  2 & 24.43 & 0.08
&  \phn0.77 & 0.03 \nl
M81      & N3031 & 147.9 &  69.3 &   \llap{$-$}40 &  2 & 27.78 & 0.08
&  \phn3.6  & 0.2 \nl
CenA     & N5128 & 200.0 & \llap{$-$}39.0 &   550 &  3 & 28.03 & 0.10
&  \phn4.0  & 0.2 \nl
N1023    & N1023 &  37.0 &  35.0 &   650 &  4 & 29.91 & 0.09 & \phn9.6
& 0.4 \nl
LeoI     & N3379 & 161.3 &  12.8 &   900 &  5 & 30.14 & 0.06 & 10.7  &
0.3 \nl
N7331    & N7331 & 338.7 &  34.2 &   800 &  2 & 30.39 & 0.10 & 12.0  &
0.6 \nl
UMa      & N3928 & 180.0 &  47.0 &   900 &  5 & 30.76 & 0.09 & 14.2  &
0.6 \nl
ComaI    & N4278 & 184.4 &  29.6 &  1000 &  3 & 30.95 & 0.08 & 15.5  &
0.6 \nl
ComaII   & N4494 & 187.2 &  26.1 &  1350 &  3 & 31.01 & 0.08 & 15.9  &
0.6 \nl
Virgo    & N4486 & 187.1 &  12.7 &  1150 & 27 & 31.03 & 0.05 & 16.1  &
0.4 \nl
Dorado   & N1549 &  63.7 & \llap{$-$}55.7 &  1300 &  6 & 31.04 & 0.06 & 16.1  &
0.5 \nl
Fornax   & N1399 &  54.1 & \llap{$-$}35.6 &  1400 & 26 & 31.23 & 0.06 & 17.6  &
0.5 \nl
\enddata
\tablecomments{Columns: 
Group name, sample member, RA and Dec (B1950), 
mean heliocentric velocity, number of SBF distances, SBF distance
modulus and error, and SBF distance (Mpc) and error.}

\end{deluxetable}

\clearpage

\figcaption[fig1.eps]{ Histograms of photometric 
differences in surface photometry between all pairs of observations
(including published photoelectric data) of the same galaxies, in $V$
(upper panel), $I$ (center panel), and \vi\ (lower panel) The points
and wider Gaussian curves are the distribution of pairwise differences
prior to application of run offsets, and the histograms and narrower
Gaussians reflect the improvement from using these offsets.  This pair
by pair comparison is used to define zero point offsets between runs
and bring all the photometry onto the same scale as photoelectric
photometry.
\label{fig1}}

\figcaption[fig2.eps]{ Differences for multiple observations of the
same galaxy in \viz\ (upper panel) and \mi\ (lower panel).  Each
histogram shows the distribution of the difference between multiple
observations divided by the estimated error.  A Gaussian of unity
variance and normalization equal to the number of pairs is shown for
comparison.
\label{fig2}}

\figcaption[fig3.eps]{ The distribution of \mi\ as a function of
\viz\ for six groups: NGC~1023, Leo, UMa, Coma~I\&II (including both the
NGC~4278 and NGC~4494 subgroups), Virgo, and Fornax.  Spiral galaxies
(i.e. RC3 T-type $T>0$) are indicated with filled symbols, and the
vertical error bar shows our estimate of rms group depth (derived from
the angular extent of the group across the sky) multiplied by 5
as an expectation for the peak-to-peak depth of the cluster.  Only in
Fornax and in Leo are the SBF measurement errors as big as the
putative depth of the group.  The three small squares in the Virgo
panel are for NGC 4600, NGC 4365, and NGC 4660, galaxies which we
believe to be foreground or background even though they meet the Virgo
group criteria.  The lines show the SBF relation for each of these groups
from our overall fit to all the data.
\label{fig3}}

\figcaption[fig4.eps]{ The distribution of \mi\ - \avemi\ as a
function of \viz\ for all galaxies belonging to our groups.  
A line through $(1.15,0.0)$
with the mean slope of the SBF-color
relation is drawn.  The four spiral galaxies for which we have both
Cepheid and SBF distances, NGC~224, NGC~3031, NGC~3368, and NGC~7331,
are plotted as large, solid hexagons and demonstrate that SBF is the same for
spiral bulges as for elliptical and S0 galaxies.
The round, solid points above the line are various locations in 
NGC~205, and the round, solid points below the line are NGC~147 and
NGC~185 (not used in the fit).  The inset shows NGC~205 again, along
with NGC~5253 and IC~4182, which are placed according to their Cepheid
distances.
\label{fig4}}

\figcaption[fig5.eps]{ The \viz\ corrected \mi\ quantity, \mim, 
as a function of Mg$_2$ and $B_T$ magnitude (from the RC3) in the
Virgo and Fornax clusters.  The lack of correlation indicates that
\mi\ is a one-parameter function of stellar population, and \viz\ 
adequately delineates the variations of stellar population over this
color range.
\label{fig5}}

\figcaption[fig6.eps]{ The mean \avemi\ derived for our groups
compared to other distance estimators:  Cepheids, SNII, SNIa (treated
as standard candles), and SNIa (correcting the peak magnitude for the
rate of decline $\Delta m_{15}$.  The other estimators' distances are
either in terms of Mpc, expressed as a distance modulus \dmod, or in
terms of km/s, plotted as 5 times the logarithm.  The lines are drawn
according to a weighted fit of unity slope between each set of
distances.  The ``fast decline'' SNIa are plotted as solid points, but
are not used in any of the fits. NGC~7331 is drawn as a solid point in
the SNII comparison because the SNII distance is discordant with both
SBF and Cepheids.  Above each distance comparison is a histogram of the
differences $\avemi-\dmod$ or $\avemi-5\log d$.
\label{fig6}}

\figcaption[fig7.eps]{ Same as Figure 6, but comparing SBF and the
tertiary estimators PNLF, TF (Mpc zero point), TF (from MarkII
catalog), and \dn.  The recent PNLF distances for the ComaI\&II galaxies
are plotted as solid symbols.
\label{fig7}}

\figcaption[fig8.eps]{ Comparison of the theoretical model
predictions of $\overline M_I$ from Worthey with our empirical
calibration from Cepheids,  $ \overline M_I = -1.74 + 4.5 [ \viz -
1.15 ]$, drawn as a dashed line.  The solid line shows a fit to the
theoretical models using the empirical slope: the two differ by 0.07
mag in zero point or 0.015 mag in color.  The models have ages of 5, 8,
12, and 17 Gyr (older are redder and fainter), and their metallicity
relative to solar is indicated by point type.
\label{fig8}}

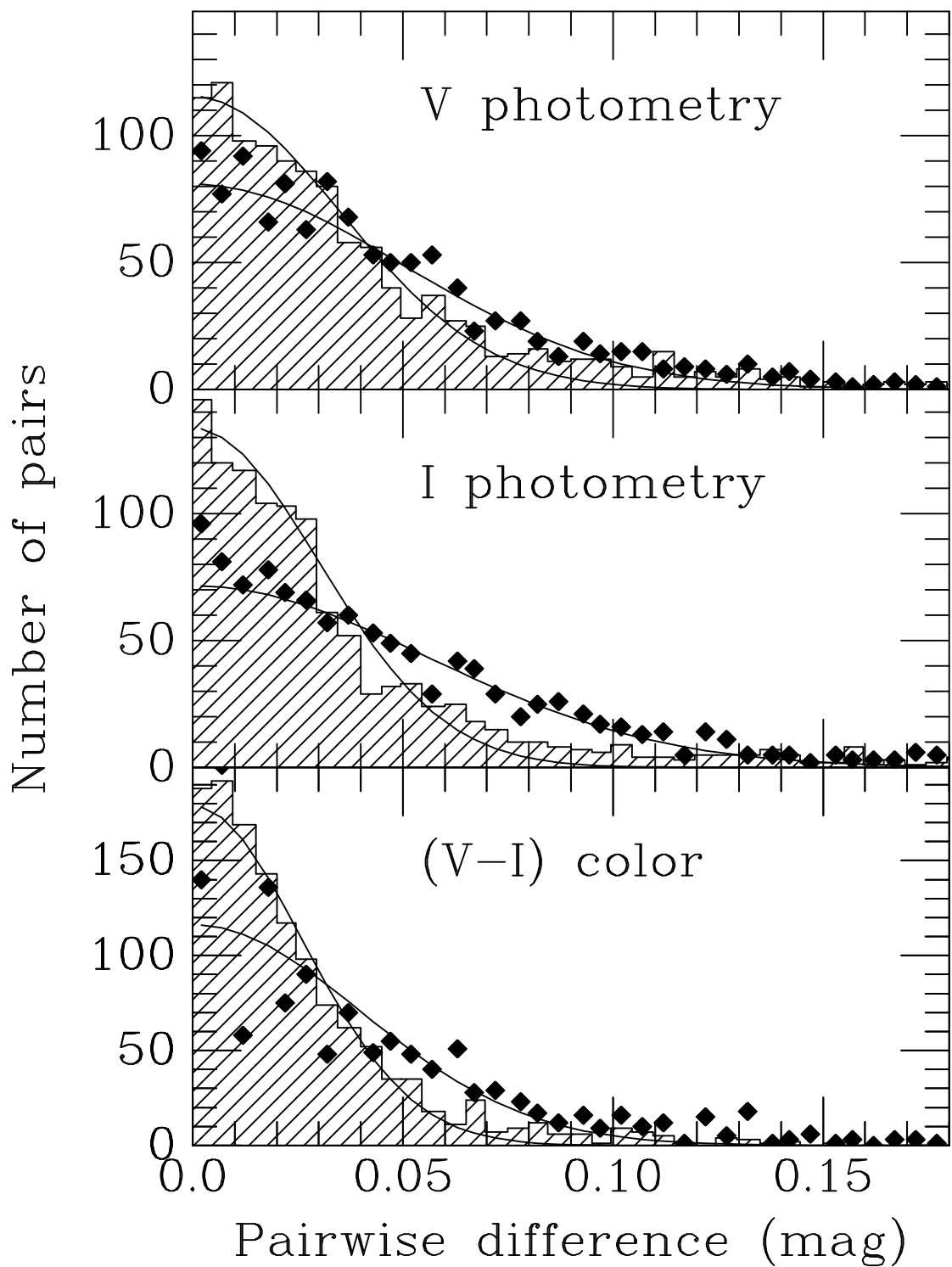
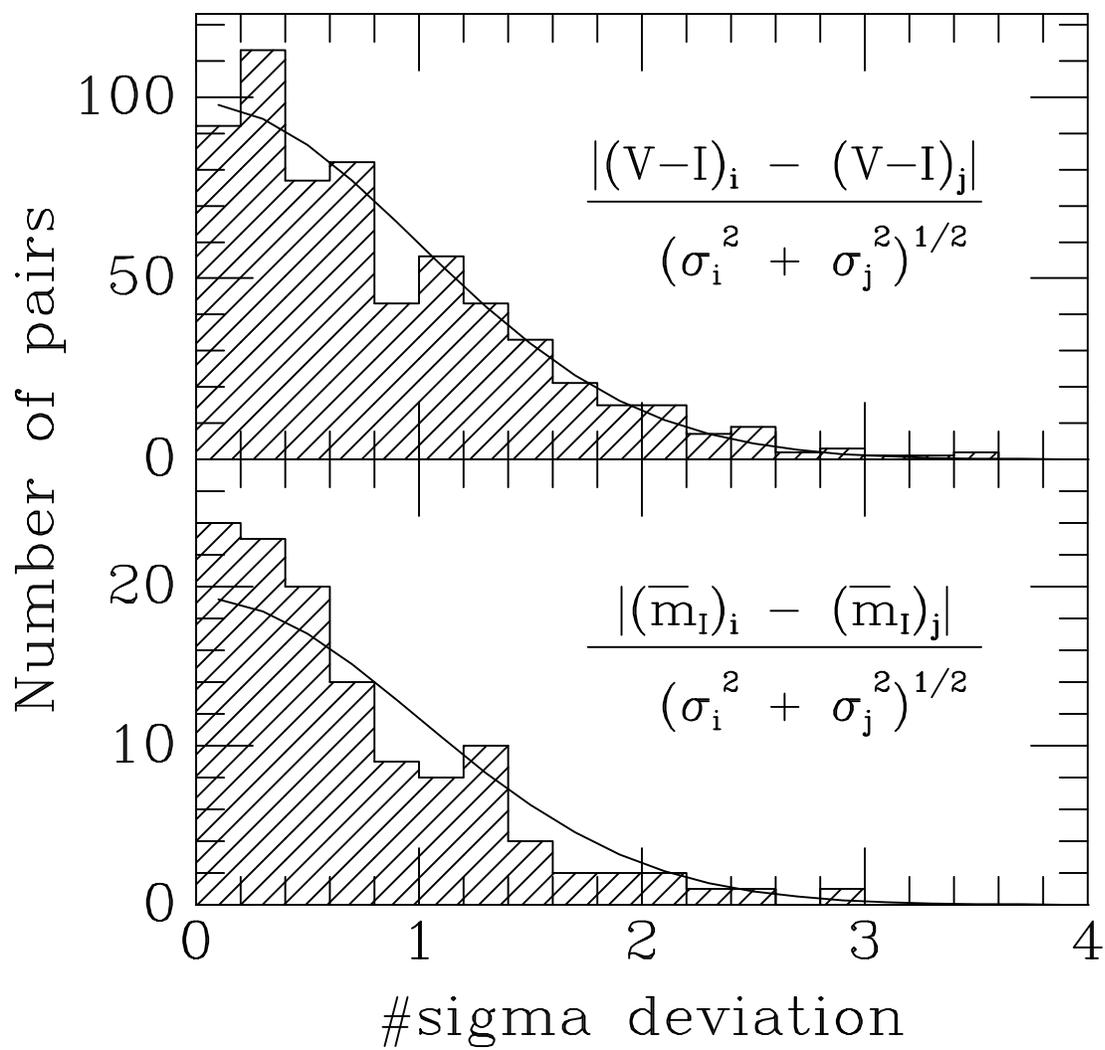
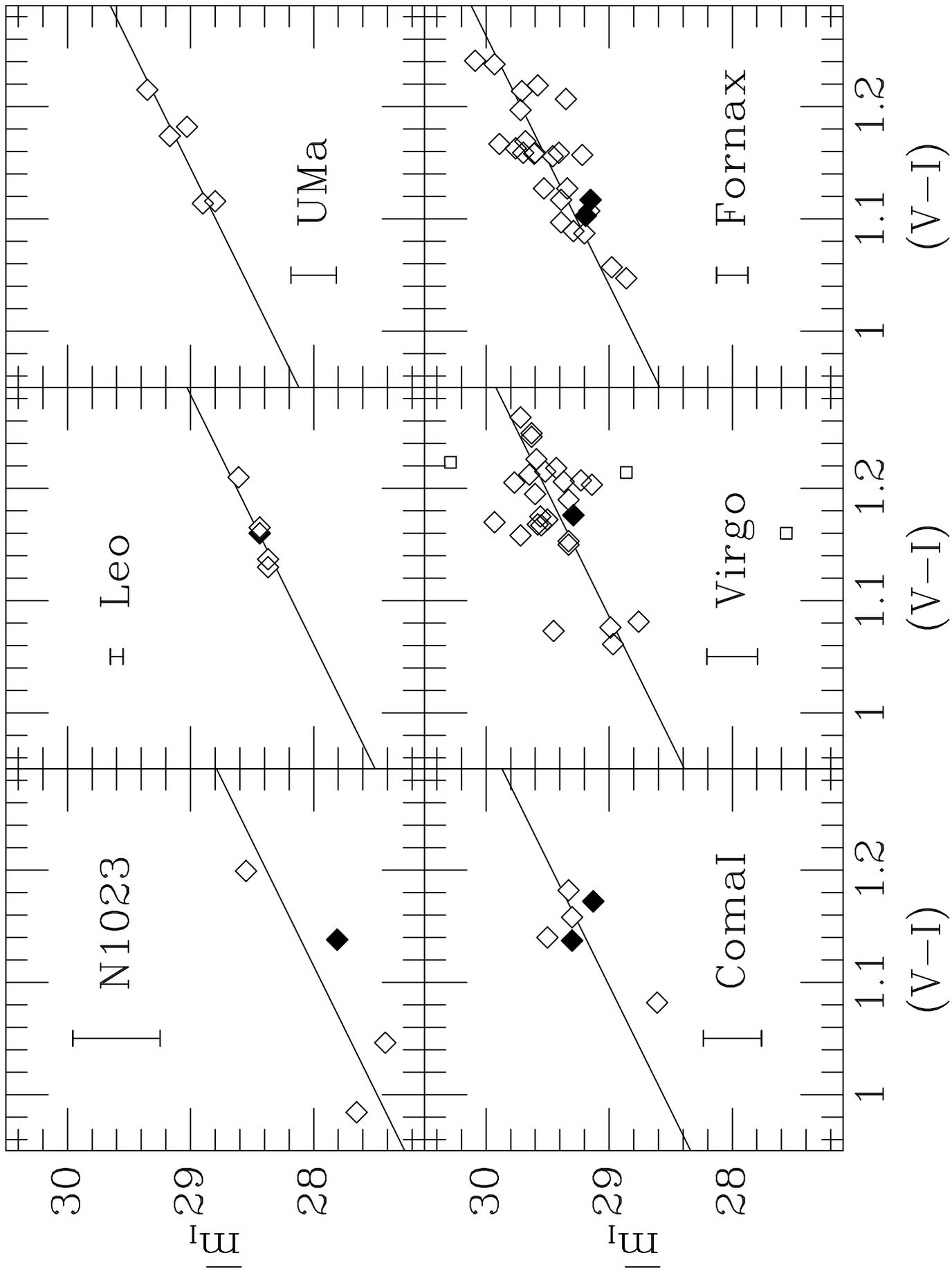
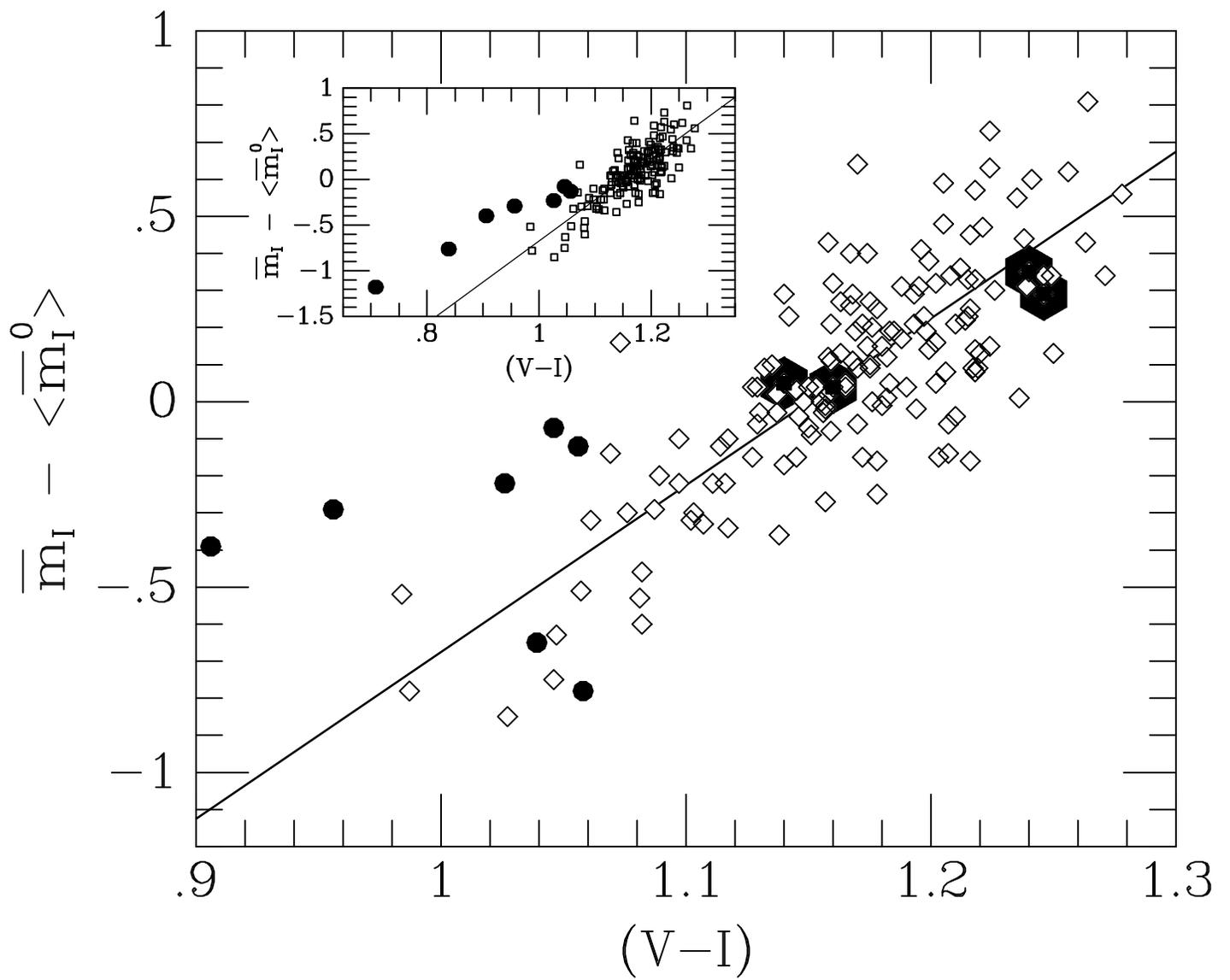
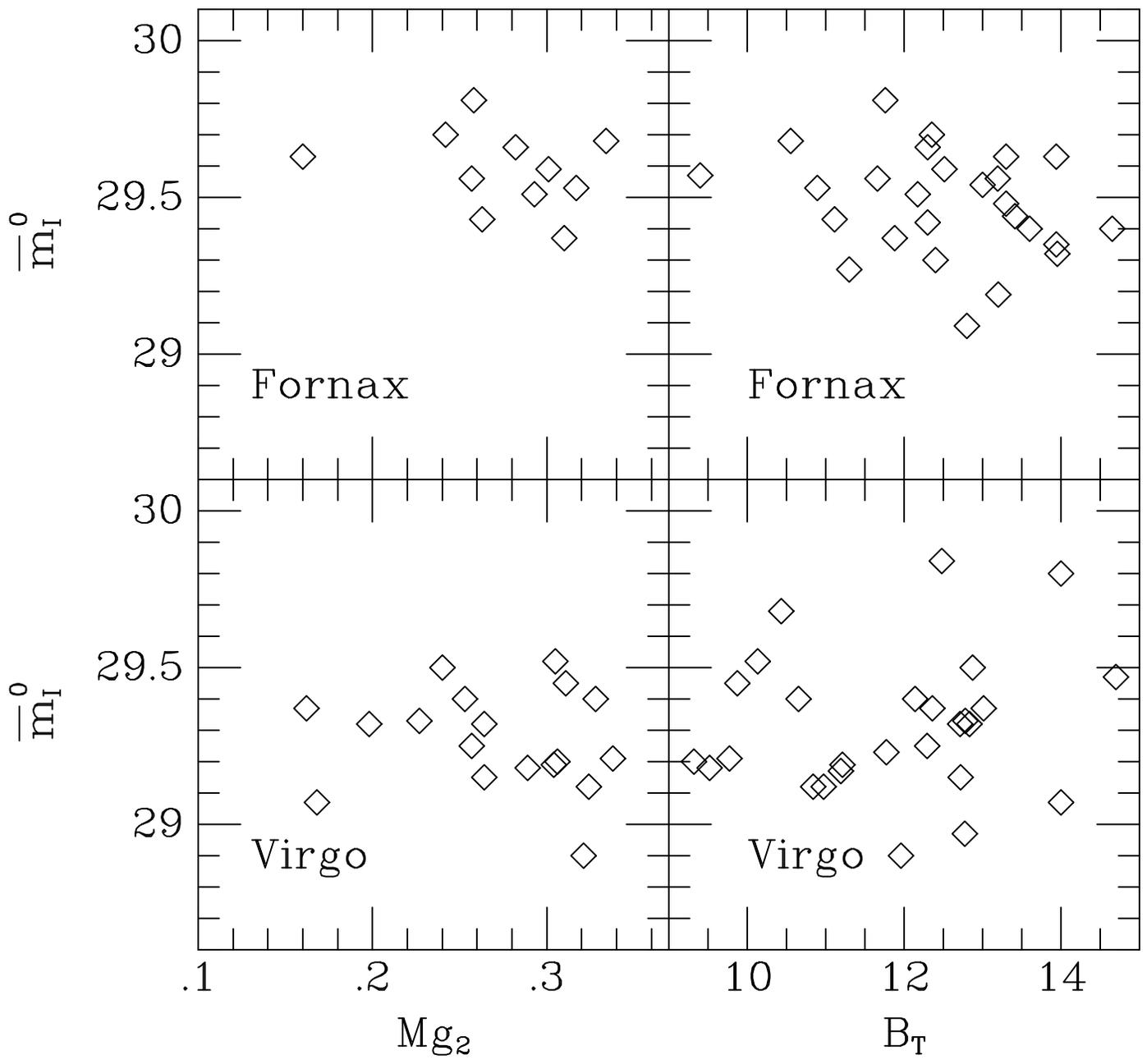
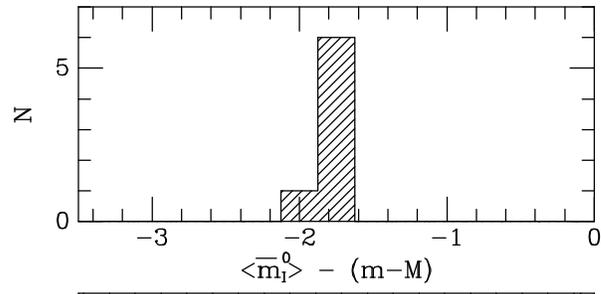
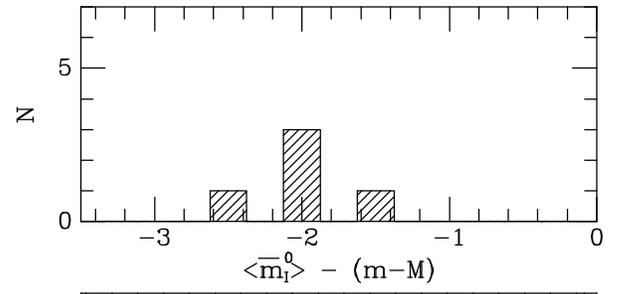
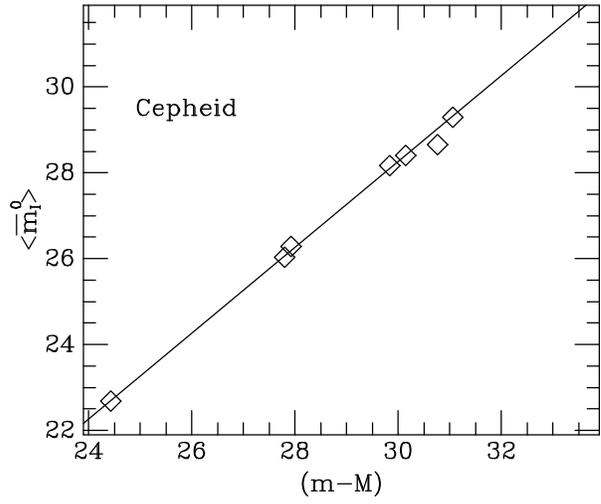
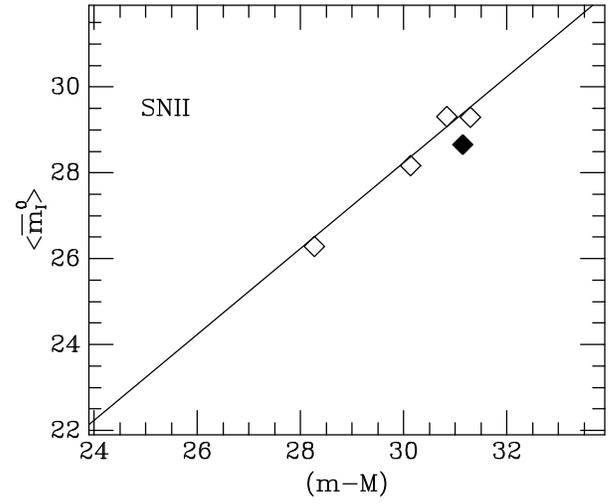
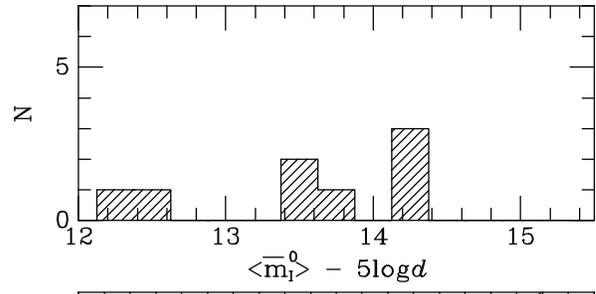
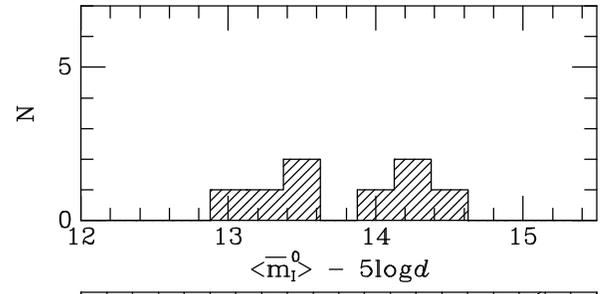
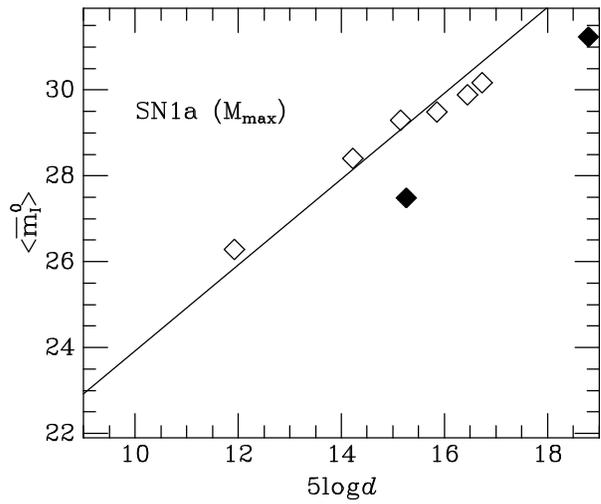
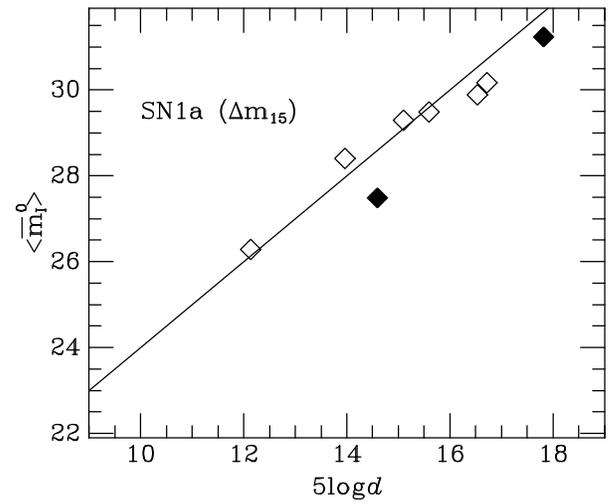
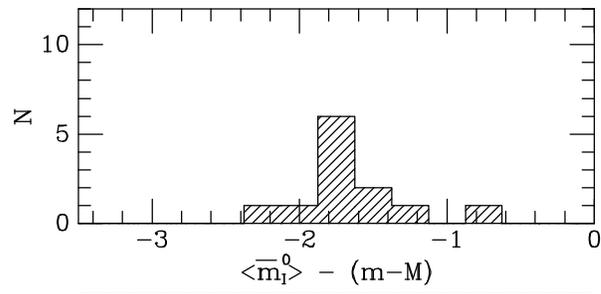
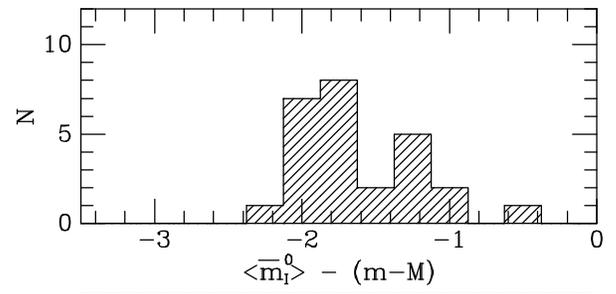
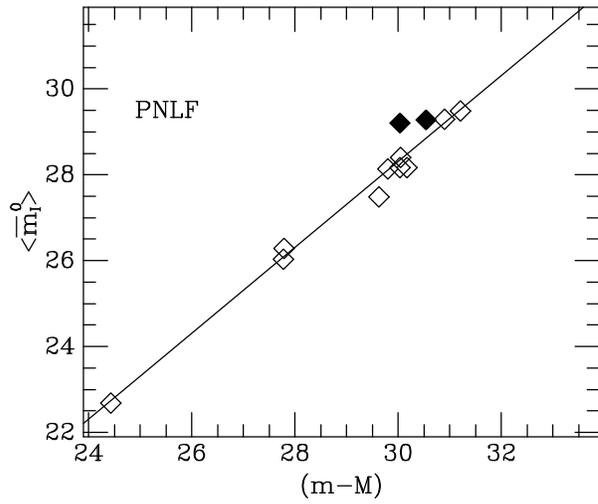
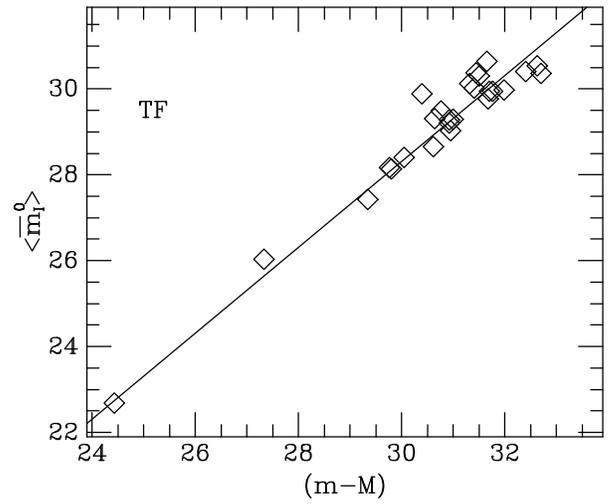
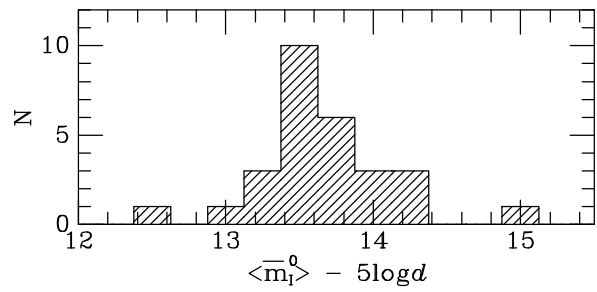
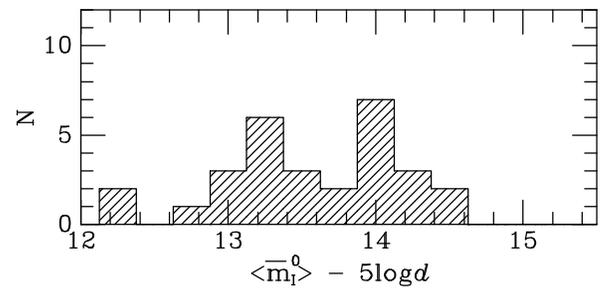
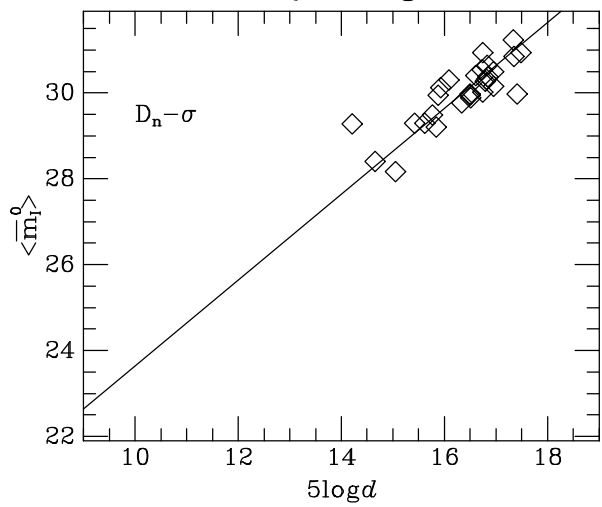
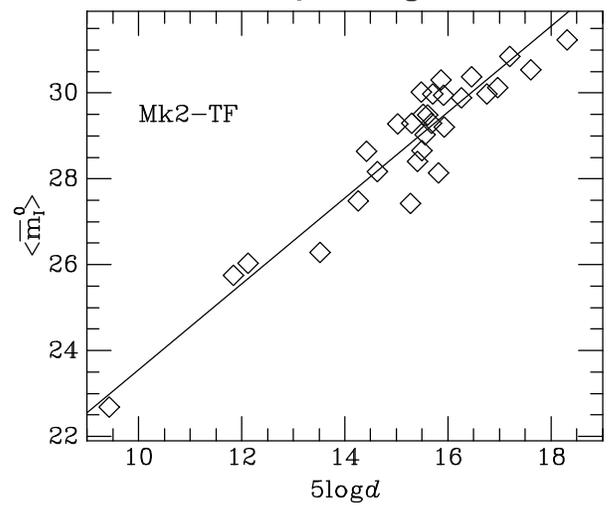
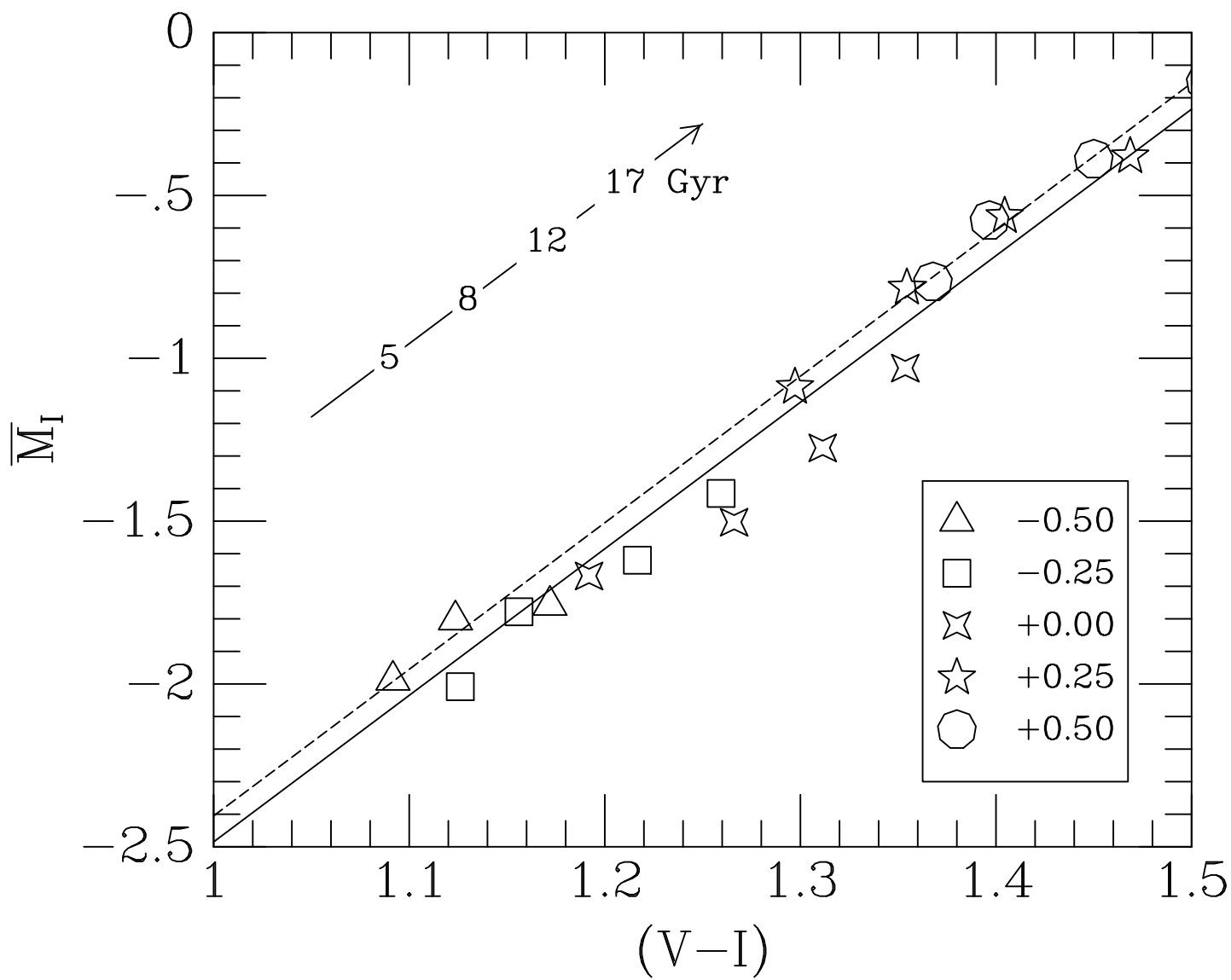
\end{document}